\documentclass[twocolumn]{aastex631}
\usepackage{multirow}

\begin{document}

\title{Ultra-deep Keck/MOSFIRE spectroscopic observations of $z\sim 2$ galaxies: direct oxygen abundances and nebular excitation properties}

\author[0000-0003-1249-6392]{Leonardo Clarke}
\affiliation{Department of Physics and Astronomy, University of California, Los Angeles, 430 Portola Plaza, Los Angeles, CA, 90095, USA}

\author[0000-0003-3509-4855]{Alice Shapley}
\affiliation{University of California, Los Angeles, 430 Portola Plaza, Los Angeles, CA, 90095, USA}

\author[0000-0003-4792-9119]{Ryan L. Sanders}
\altaffiliation{NHFP Hubble Fellow}
\affiliation{University of California, Davis, One Shields Ave, Davis, CA 95616, USA}

\author[0000-0001-8426-1141]{Michael W. Topping}
\affiliation{Steward Observatory, University of Arizona, 933 N Cherry Avenue, Tucson, AZ 85721, USA}

\author[0000-0001-5860-3419]{Tucker Jones}
\affiliation{University of California, Davis, One Shields Ave, Davis, CA 95616, USA}

\author[0000-0002-7613-9872]{Mariska Kriek}
\affiliation{Leiden Observatory, Leiden University, P.O. Box 9513, NL-2300 AA Leiden, The Netherlands}

\author[0000-0001-9687-4973]{Naveen A. Reddy}
\affiliation{Department of Physics \& Astronomy, University of California, Riverside, 900 University Avenue, Riverside, CA 92521, USA}

\author{Daniel P. Stark}
\affiliation{Steward Observatory, University of Arizona, 933 N Cherry Avenue, Tucson, AZ 85721, USA}

\author[0000-0001-5940-338X]{Mengtao Tang}
\affiliation{Steward Observatory, University of Arizona, 933 N Cherry Avenue, Tucson, AZ 85721, USA}



\newcommand{\oiii}{[O\thinspace{\sc iii}]}
\newcommand{\neiii}{[Ne\thinspace{\sc iii}]}
\newcommand{\oii}{[O\thinspace{\sc ii}]}
\newcommand{\hii}{H\thinspace{\sc ii} }
\newcommand{\nii}{[N\thinspace{\sc ii}]}
\newcommand{\niii}{[N\thinspace{\sc iii}]}
\newcommand{\feii}{[Fe\thinspace{\sc ii}]}
\newcommand{\sii}{[S\thinspace{\sc ii}]}
\newcommand{\oi}{[O\thinspace{\sc i}]}
\newcommand{\secpoint}{\mbox{$''\mskip-7.6mu.\,$}}

\begin{abstract}

Using deep near-infrared Keck/MOSFIRE observations, we analyze the rest-optical spectra of eight star-forming galaxies in the COSMOS and GOODS-N fields. We reach integration times of $\sim$10 hours in the deepest bands, pushing the limits on current ground-based observational capabilities. The targets fall into two redshift bins --- 5 galaxies at $z \sim 1.7$ and 3 at $z \sim 2.5$ --- and were selected as likely to yield significant auroral-line detections. Even with long integration times, detection of the auroral lines remains challenging. We stack the spectra together into subsets based on redshift, improving the signal-to-noise ratio on the \oiii $\lambda 4364$ auroral emission line and, in turn, enabling a direct measurement of the oxygen abundance for each stack. We compare these measurements to commonly-employed strong-line ratios alongside measurements from the literature. We find that the stacks fall within the distribution of $z>1$ literature measurements, but a larger sample size is needed to robustly constrain the relationships between strong-line ratios and oxygen abundance at high redshift. We additionally report detections of \oi $\lambda6302$ for nine individual galaxies and composite spectra of 21 targets in the MOSFIRE pointings. We plot their line ratios on the \oiii $\lambda 5008$/H$\beta$ vs. \oi $\lambda 6302$/H$\alpha$ diagnostic BPT diagram, comparing our targets to local galaxies and \hii regions. We find that the \oi/H$\alpha$ ratios in our sample of galaxies are consistent with being produced in gas ionized by $\alpha$-enhanced massive stars, as has been previously inferred for rapidly-forming galaxies at early cosmic times.

\end{abstract}



\section{Introduction} \label{sec:intro}

Tracing the chemical evolution of galaxies is key to understanding how galaxy growth and evolution occur over time. The metallicity of a galaxy is influenced by numerous mechanisms such as the reprocessing of gas into heavier elements through nucleosynthesis; metal-enriched outflows driven by supernovae, AGN, and stellar winds; accretion of pristine hydrogen gas onto a galaxy; and accretion of enriched, recycled gas in the form of galactic fountains \citep{2017ARA&A..55..389T,2017MNRAS.467..115D}. Observationally, metallicity commonly refers to the gas-phase oxygen abundance in the interstellar medium (ISM) of a galaxy since oxygen is the most abundant metal and produces strong rest-optical emission line features. The oxygen abundance in the ISM of star-forming galaxies has been observed to correlate tightly with the stellar mass, encapsulated in what is referred to as the mass-metallicity relation (MZR). Early evidence for a MZR goes back to \citet{1979A&A....80..155L} who measured oxygen abundances in a small sample of nearby blue compact dwarf galaxies. Later studies \citep[e.g.,][]{2004ApJ...613..898T,2008ApJ...681.1183K,2013ApJ...765..140A} showed that there is a MZR that generally describes galaxies in the local universe. Furthermore, many works \citep[e.g.,][]{2006ApJ...644..813E,2008A&A...488..463M,2009MNRAS.398.1915M,2011ApJ...730..137Z,2014ApJ...795..165S,2014ApJ...791..130Z,2015PASJ...67..102Y,2016ApJ...828...67L,2016ApJ...822..103G,2021ApJ...914...19S} have revealed an evolution in the MZR with redshift, noting a change in the turnover mass and the normalization at higher $z$. Folding in the global star-formation rate (SFR) to the MZR yields the fundamental metallicity relation (FMR). This relation appears not to evolve through cosmic time at least as far back as $z\sim 3$ \citep[e.g.,][]{2010MNRAS.408.2115M,2021ApJ...914...19S,2022arXiv221206877H}, though there is evidence of deviations from the FMR at higher redshifts \citep[e.g.,][]{2023arXiv230408516C}.

The existence of these scaling relations with galaxy parameters gives insight into the processes that govern galaxy formation. The SFR, which is governed by the gas reservoir in a galaxy, is influenced by the baryon cycle, and is therefore tied to the chemical evolution in the ISM through the processes described above. Additionally, the stellar mass represents the integrated sum of star formation and is also related to the total metal production across a galaxy's lifetime. Overall, the three parameters that comprise the FMR probe the important mechanisms that determine galaxy evolution. The invariance of the FMR through a large portion of cosmic history suggests that galaxies are driven towards an equilibrium among inflow, star formation, and outflow \citep[e.g.,][]{2012MNRAS.421...98D, 2014MNRAS.443.3643P}. These scaling relations are additionally useful for hydrodynamical simulations of galaxy formation since the comparison with observations provides further constraints on the subgrid physics determining the outputs \citep[e.g.,][]{2016MNRAS.456.2140M,2017MNRAS.467..115D,2017MNRAS.472.3354D, 2019MNRAS.484.5587T}. Thus, measuring these galaxy parameters with high accuracy is instrumental in our understanding of galaxy formation and evolution.

Making robust measurements of the metallicity of a galaxy, however, can prove quite challenging. One of the more physically-motivated methods of determining the gas-phase oxygen abundance of a galaxy involves measuring the average electron temperature and density of its \hii regions. From these physical properties, one can determine the emissivities of the emission-line transitions from each ion species which, when scaled by their respective line flux measurements, yields the abundance of each ion relative to hydrogen \citep[i.e. O$^+$/H$^+$ and O$^{2+}$/H$^+$, see][for more detail]{2006A&A...448..955I,2015A&A...573A..42L,2017PASP..129h2001P}. This method of abundance determination is often referred to as the ``direct" method. However, to obtain the electron temperature, one must be able to detect a set of faint rest-optical auroral emission lines (e.g. \oiii $\lambda 4364$, \oii $\lambda\lambda 7322,7332$), which can be a hundred times fainter than their nebular counterparts, requiring very long exposure times \citep{1992AJ....103.1330G, 2017PASP..129d3001P}. The task becomes increasingly challenging when observing targets at $z>1$ due to the varying transmission of Earth's atmosphere in the near-infrared (i.e., rest-frame optical) and the apparent faintness of targets due to their increased distance. Additionally, in the case of the \oiii $\lambda 4364$ line, whose strength relative to \oiii $\lambda 5008$ is temperature-sensitive, detection becomes more difficult in higher-metallicity galaxies. This challenge is due to the effects of more efficient metal-line cooling which leads to lower electron temperatures in the constituent \hii regions in more metal-rich galaxies, rendering direct metallicity measurements much more difficult.

In light of the challenges associated with the direct method, it is common to determine metal abundances using indirect metallicity indicators which rely on the ratios of strong emission lines \citep[e.g.,][]{2013ApJS..208...10D, 2016MNRAS.457.3678P,2018ApJ...859..175B,2020MNRAS.491..944C,2022ApJS..262....3N}. These relations translating strong-line ratios to metallicity are calibrated to photoionization models and/or measurements of direct metallicity in \hii regions and galaxies in the local universe. It is uncertain, however, whether these indirect indicators remain accurate at higher redshifts since conditions in the ISM of galaxies at earlier cosmic times may not resemble those in the local universe. Current observations suggest that galaxies at $z\sim 2$ may be characterized by harder ionizing spectra and may also have N/O ratios that vary slightly from galaxies in the local universe at fixed oxygen abundance \citep[e.g.,][]{2015ApJ...801...88S,2017ApJ...836..164S,2019ApJ...881L..35S,2022arXiv221206877H}. 

The evolving conditions in the ISM of galaxies has typically been traced by measurements of both high- and low-ionization emission lines (e.g., \oiii/H$\beta$, \nii/H$\alpha$, \sii/H$\alpha$). However, one low-ionization diagnostic that has been missing in analyses of galaxies at $z>1$ is the \oi $\lambda 6302$/H$\alpha$ line ratio. In studies of low-redshift galaxies, this line ratio offers insights into the hardness of the ionizing spectrum, the contribution of diffuse ionized gas (DIG), and the presence of shocks \citep[e.g.,][]{2017MNRAS.466.3217Z}. However, due to the intrinsic faintness of the \oi $\lambda 6302$ line, studies of this line diagnostic at high redshift have typically proven very difficult. 

Similarly, due to the difficulty of obtaining auroral-line measurements at high redshift with ground-based facilities, there only exists a small sample of $z>1$ galaxies in the literature for which direct oxygen abundances have been measured \citep[examples from the literature are discussed in][]{2020MNRAS.491.1427S}. Additionally, the integration times for most ground-based near-infrared spectroscopic surveys do not reach the required depth to achieve significant auroral-line detections. The prevalence of auroral-line detections is, however, increasing as a result of observations made by the new \textit{James Webb Space Telescope} (JWST) \citep[e.g.,][]{2023MNRAS.518..425C, 2023arXiv230112825N, 2023arXiv230308149S}. 

The deep spectral observations analyzed in this study push the limits of ground-based, 10-m-class observatories and highlight the importance of JWST and other future state-of-the-art observatories in characterizing galaxy properties at and beyond cosmic noon. In this study, we utilize the MOSFIRE instrument \citep{2012SPIE.8446E..0JM} on the 10-m Keck I telescope to make deep observations of a sample of eight galaxies at $z\sim 1.7-2.5$, reaching up to $\sim$10 hours of integration time in some bands. We additionally produce composite spectra from this sample of eight galaxies and determine their average characteristics (i.e., chemical abundances, electron temperatures, and densities). The depth of these observations additionally enables the detection of the \oi $\lambda 6302$ feature, beyond the reach of more typical spectroscopic samples with shallower integration times \citep[e.g.,][]{2015ApJS..218...15K}. Based on these measurements, we investigate the position of our sample of $z\sim 2$ galaxies in the \oiii$\lambda 5008$/H$\beta$ vs. \nii$\lambda 6585$/H$\alpha$, \sii $\lambda \lambda 6718,6733$/H$\alpha$, and \oi$\lambda 6302$/H$\alpha$ diagnostic diagrams (hereafter BPT diagrams\footnote{BPT refers to \citet{1981PASP...93....5B}, though the \oiii$\lambda 5008$/H$\beta$ vs. \sii$\lambda\lambda 6716, 6731$/H$\alpha$ diagnostic diagram was later introduced by \citet{1987ApJS...63..295V}.}), the latter probing an unexplored parameter space at $z>1$.

In Section \ref{sec:methods} of this paper, we give an overview of the observational setup and data processing methods. In Section \ref{sec:results}, we present the results of our analysis of the spectra in our sample. In Section \ref{sec:discussion}, we compare our measurements with commonly-used metallicity indicators as well as consider the nature of \oi $\lambda 6302$ emission at high redshift. Throughout, we adopt the following cosmological parameters: $H_0=70$ km s$^{-1}$ Mpc$^{-1}$, $\Omega_{\rm m}=0.3$, and $\Omega_{\Lambda}=0.7$. Additionally, we assume a \citet{2003PASP..115..763C} IMF, solar abundances of $\rm 12+\log(O/H)_{\odot}=8.69$ and $\rm 12+\log(N/H)_{\odot}=7.83$, and a solar metallicity of $Z_{\odot}=0.014$ \citep{2009ARA&A..47..481A}.

\begin{deluxetable*}{lccccccc}
\tabletypesize{\footnotesize}
\caption{\oiii\ auroral targets and physical properties.}
\label{tab:galaxies}
\tablehead{
\colhead{ID} & \colhead{R.A.} & \colhead{Dec.} & \colhead{$z$} 
 & \colhead{$\log{(M_*/M_{\odot})}$} & \colhead{$\log{(t_{age}/yr)}$} & \colhead{SFR} & \colhead{sSFR}\\
\colhead{} & \colhead{(J2000)} & \colhead{(J2000)} & \colhead{} & \colhead{} & \colhead{} & \colhead{$(M_{\odot}~yr^{-1})$} & \colhead{$(Gyr^{-1})$}
}
\startdata
COSMOS-18812  & 10:00:36.896 & +02:22:13.82 & 2.46236 & $8.74^{+0.07}_{-0.04}$  & $8.30^{+0.14}_{-0.18}$  & $7.47^{+4.26}_{-2.53}$     & $13.26 \pm 6.42$\\
COSMOS-19439  & 10:00:24.360 & +02:22:36.20 & 2.46598 & $10.26^{+0.00}_{-0.00}$ & $9.40^{+0.00}_{-0.00}$  & $141.46^{+48.18}_{-34.46}$ & $7.90 \pm 2.25$\\
COSMOS-19753  & 10:00:18.182 & +02:22:50.31 & 2.46884 & $10.55^{+0.04}_{-0.00}$ & $9.40^{+0.00}_{-0.06}$  & $68.89^{+4.82}_{-4.54}$    & $1.94 \pm 0.16$\\
GOODS-N-6699  & 12:36:23.385 & +62:10:29.04 & 1.66448 & $9.82^{+0.05}_{-0.05}$  & $9.50^{+0.00}_{-0.40}$  & $11.51^{+2.62}_{-2.20}$    & $1.78 \pm 0.42$\\
GOODS-N-8013  & 12:36:52.008 & +62:10:54.80 & 1.66776 & $9.44^{+0.04}_{-0.03}$  & $8.80^{+0.10}_{-0.09}$  & $6.68^{+0.70}_{-0.69}$     & $2.43 \pm 0.31$\\
GOODS-N-8240  & 12:36:25.249 & +62:10:58.91 & 1.69090 & $9.76^{+0.02}_{-0.13}$  & $9.20^{+0.18}_{-0.60}$  & $12.36^{+9.65}_{-6.22}$    & $2.14 \pm 1.43$\\
GOODS-N-14595 & 12:36:13.373 & +62:12:49.91 & 1.67596 & $9.02^{+0.09}_{-0.11}$  & $8.90^{+0.20}_{-0.50}$  & $9.60^{+3.36}_{-2.29}$     & $9.08 \pm 3.61$\\
GOODS-N-18462 & 12:36:11.906 & +62:13:58.80 & 1.67463 & $9.52^{+0.05}_{-0.03}$  & $9.30^{+0.11}_{-0.10}$  & $3.46^{+0.30}_{-0.29}$     & $1.04 \pm 0.13$\\
\enddata
\tablecomments{Some of the uncertainties on the stellar masses and ages output from FAST are quoted as being $\pm0.00$, which only represents an uncertainty on the fitting of the models to the data. It does not account for systematic uncertainties, which were estimated to be $\sim$0.1 dex in a similar analysis by \citet{2009ApJ...701.1839M}.}
\end{deluxetable*}

\section{Methods and Observations} \label{sec:methods}

\subsection{Sample Selection and Observation Configurations}

The target galaxies in this analysis reside in the COSMOS and GOODS-N extragalactic legacy fields covered by the CANDELS and 3D-HST surveys \citep{2011ApJS..197...35G,2011ApJS..197...36K,2016ApJS..225...27M}. These galaxies were drawn from the MOSFIRE Deep Evolution Field (MOSDEF) survey \citep{2015ApJS..218...15K} as well as the sample of extreme emission-line galaxies (EELGs) presented by \citet{2019MNRAS.489.2572T} selected from the 3D-HST WFC3 grism emission-line catalog \citep{2016ApJS..225...27M}. Targets of interest were selected based on their probability of yielding auroral-line detections. To this effect, the target sample was composed of galaxies that had bright \oiii $\lambda 5008$ emission and whose ratios among strong nebular emission lines suggested a high electron temperature based on relations observed in $z\sim 0$ \hii regions \citep{2017ApJ...850..136S}. Such thermal properties would result in brighter \oiii $\lambda 4364$ and \oii $\lambda\lambda 7322,7332$ auroral lines, necessary for making direct metallicity measurements. We also required galaxy targets to lie within the following redshift intervals: $1.62 \leq z \leq 1.70$, $2.32 \leq z \leq 2.61$, and $2.95 \leq z \leq 3.18$ to capture both auroral and strong nebular emission lines in the near-infrared atmospheric transmission windows covered by the Y, J, H, and K-bands.

In light of these selection criteria, we targeted a sample of 12 galaxies, which we refer to as ``auroral" targets. Seven of the auroral targets were in the COSMOS field and five were in the GOODS-N field. Two of the galaxies from the COSMOS field are the subjects of a recent paper by \citet{2023ApJ...943...75S} in which their oxygen abundances were measured via the \oii $\lambda\lambda 7322, 7332$ auroral line doublet, representing the first such measurements beyond the local universe.

From the remaining sample of 10 \oiii\ auroral targets, two from the COSMOS field were not considered in this study. The first of these targets was excluded because the galaxy was dithered on top of an adjacent object in the field, causing the target signal to be strongly contaminated by the negative trace of its neighbor on the slit. The second of these targets was at $z=3.12$, placing H$\alpha$ beyond the coverage of the K-band, and the remaining higher-order Balmer emission lines fell onto atmospheric sky lines. As a result, it was not possible to apply Balmer-decrement-based dust corrections on this particular target. Because of these considerations, our final auroral-target sample consisted of eight galaxies: three at $z\sim 2.5$ in the COSMOS field and five at $z\sim 1.7$ in the GOODS-N field. The properties of these targets are summarized in Table \ref{tab:galaxies}. We utilized the multiplexing capabilities of MOSFIRE to observe an additional 29 filler targets (14 in the COSMOS pointing and 15 in the GOODS-N pointing). For the subset of 15 filler-target spectra that contained H$\alpha$, H$\beta$, and \oi $\lambda 6302$ coverage, spanning $z=1.405$ to $z=2.515$, we analyzed them for \oi $\lambda 6302$ emission, and we refer to these targets as ``non-auroral" targets.

Observations were collected over six nights: January 13, 2019 (COSMOS $H$-band); March 16, 2019 (COSMOS $J$- and $H$-bands); March 3, 2021 (GOODS-N $J$-band); March 4, 2021 (COSMOS $K$-band, GOODS-N $J$-band), April 19, 2021 (GOODS-N $Y$-, $J$-, and $H$-bands), and May 1, 2021 (GOODS-N $H$-band). The observations were taken with 0\secpoint7 slits using an ABA'B' dither pattern, and dithered frames were aligned and combined to perform sky subtraction. A summary of the observation configurations and conditions can be found in Table \ref{tab:obs}.

\begin{deluxetable}{l|cccccc}
    \caption{Observation specifications for each band and pointing.}
    \label{tab:obs}
    \tablehead{
     Pointing & \multicolumn{3}{c|}{COSMOS} & \multicolumn{3}{c}{GOODS-N}\\
     \cline{1-1} Band & \colhead{J} & \colhead{H} & \colhead{K} \vline & \colhead{Y} & \colhead{J} & \colhead{H}
    }
    \startdata
        Single exposure (sec) & 120 & 120 & 180 & 180 & 120 & 120\\
        Total integration (hr) & 1.86 & 8.05 & 3.88 & 1.39 & 9.71 & 1.06\\
        Spectral resolution \textit{R} & 3000 & 3650 & 3600 & 3400 & 3000 & 3650\\
        Median seeing & 0\secpoint79 & 0\secpoint53 & 0\secpoint47 & 0\secpoint84 & 0\secpoint61 & 0\secpoint67\\
    \enddata
    \tablecomments{For COSMOS-19439 and COSMOS-19753, there were existing observations from the MOSDEF survey with 2 hours in each band. We therefore combined the deep MOSFIRE spectra with existing MOSDEF observations for these two targets, thereby increasing the total exposure time by 2 hours.}
\end{deluxetable}

\subsection{Data Reduction and Flux Calibration}

The two-dimensional (2D) reduction of the MOSFIRE data was performed using an IDL data processing pipeline described in \citet{2015ApJS..218...15K}. The extractions of the 2D spectra were accomplished for each target on each mask individually using the \texttt{bmep}\footnote{https://github.com/billfreeman44/bmep} \citep{2019ApJ...873..102F} IDL program, and we used the same slit-loss correction routine as in \citet{2015ApJ...806..259R} and \citet{2015ApJS..218...15K}.

In order to monitor seeing conditions and carefully combine individual frames, a slit was placed on a star in each mask. Each exposure was weighted according to the observed flux of the slit star. An issue arose with the flux calibrations from the targets on the COSMOS mask due to the fact that a galaxy in the field was dithered on top of the slit star for that mask, thereby contaminating the slit star spectrum that was used to apply the absolute flux scaling. Consequently, the default flux calibration for the COSMOS mask was unreliable. This effect had varying significance in each of the bands. In order to mitigate this source of systematic error and to ensure accurate band-to-band flux calibrations, we compared the spectra with available photometric observations and with corresponding spectral observations from the MOSDEF survey. The existing MOSDEF spectra were taken with different slit mask configurations from those used for the new, deep observations and thus do not suffer from the same dithering issue.

The targets in the COSMOS field that had corresponding data in the MOSDEF survey were scaled multiplicatively such that the total flux of significantly-detected lines ($>$5$\sigma$) matched the flux of the same lines in the MOSDEF spectra. For targets on the COSMOS mask with no corresponding MOSDEF observations or without $>$5$\sigma$ line detections in both data sets, the spectra in each band were scaled by the average of the scaling factors on the mask in each respective filter. On average, these scaling factors adjusted the flux calibrations in each band on the order of 3-30\%.

Additionally, the spectra on the GOODS-N mask were compared with existing observations in order to ensure the robustness of the flux calibrations. Since the targets on this mask did not have corresponding MOSDEF observations, the spectra in each band were scaled to agree with broad-band ground-based photometry as well as 3D-HST \citep{2014ApJS..214...24S,2016ApJS..225...27M} photometric and spectroscopic measurements. Based on 11 galaxies with both MOSDEF coverage and photometric measurements, we find that the median photometric scaling is a factor of 1.2 higher than the scalings from the MOSDEF spectra. However, we prioritize the MOSDEF-spectra-based scalings when available, due to the better match in wavelength coverage compared to the photometric bandpasses as well as consistency with previous methodologies \citep[see][]{2023ApJ...943...75S}

\subsection{SED and Emission-line Fitting}

The spectral energy distributions (SEDs) of each target galaxy in the COSMOS field were fit across 43 photometric data points drawn from the 3D-HST catalog spanning from 3500 \AA\ to 8 $\micron$ in the observed frame. Similarly, for the GOODS-N targets, 22 photometric points were fit across the same wavelength range. We corrected the near-IR photometric data for bright rest-frame optical emission lines using the emission-line fluxes determined from the MOSFIRE spectra analyzed in this study. The SEDs were fit with flexible stellar population synthesis models \citep{2010ascl.soft10043C} using the FAST fitting code \citep{2009ApJ...700..221K} in order to determine parameters such as stellar mass and age. We assumed an SMC attenuation curve with a stellar metallicity of 0.22 $Z_{\odot}$, a delayed-$\tau$ star-formation history of the form $t \times \exp{(-t/\tau)}$, and a \citet{2003PASP..115..763C} IMF.

We used the non-linear least squares algorithm \texttt{scipy.curve\_fit()} to fit a Gaussian profile to the emission line features in each of the individual spectra as well as the stacks. Uncertainties on the line flux measurements were determined using a Monte-Carlo simulation in which each spectrum was perturbed according to the error spectrum over 100 iterations. The weaker emission-line widths were tied to the velocity widths of H$\alpha$ and \oiii $\lambda5008$. Additionally, the H$\alpha$ and \nii\ lines were fit simultaneously, with the fluxes of \nii $\lambda 6550$ and \nii $\lambda 6585$ being tied in a ratio of 1:3 respectively. Similarly to the \nii\ lines, the \oi $\lambda 6302$ and \oi $\lambda 6365$ lines were fixed with a flux ratio of 3:1 respectively since, in both cases, their relative strengths are fixed by quantum mechanical transition probabilities. For targets that appeared to have a broad/offset component to their emission-line profiles (e.g., non-auroral targets COSMOS-19812, 19985, 20062), we fit the brightest lines with a double Gaussian and reported only the narrow-component flux since the additional component is likely attributed to outflows or other gas not physically associated with \hii regions \citep{2017ApJ...849...48L}. The SED fits determined from the photometry were used to model the continuum and the Balmer stellar absorption troughs. The initial, non-emission-line corrected SEDs were used to obtain a continuum fit and estimate line fluxes, and these line flux estimates were then used to correct the near-IR photometry. In turn, we re-fit the emission lines using the corrected SEDs to obtain our final line flux measurements. These line fluxes are reported along with derived physical quantities in Table \ref{tab:all_flux}.

\subsection{Composite spectra}

We found that the signal-to-noise (S/N) on \oiii $\lambda 4364$ was very low ($<2$$\sigma$) across most of the sample, so we created composite spectra (or ``stacks") in order to boost S/N and derive average galaxy properties for each stack. We chose to stack our sample using three different configurations that are laid out in Table \ref{tab:stacked_flux}. Stack 1 (S1), consisted of the three $z\sim 2.5$ \oiii\ auroral targets in the COSMOS field, stack 2 (S2) consisted of the five $z\sim 1.7$ GOODS-N auroral targets, and stack 3 (S3) consisted of all \oiii\ auroral targets in this study. In addition to the \oiii\ auroral-line stacks, we also created composite spectra that consisted of both auroral and filler targets in the MOSFIRE pointings that had coverage of the \oi $\lambda 6302$ line unaffected by atmospheric sky lines. We divided these ``\oi" stacks into two groups: a low-redshift $(1\leq z < 2)$ stack consisting of 13 targets, and a high-redshift $(2 \leq z \leq 3)$ stack consisting of 8 targets.

\begin{figure*}
    \centering
    \includegraphics[width=17cm]{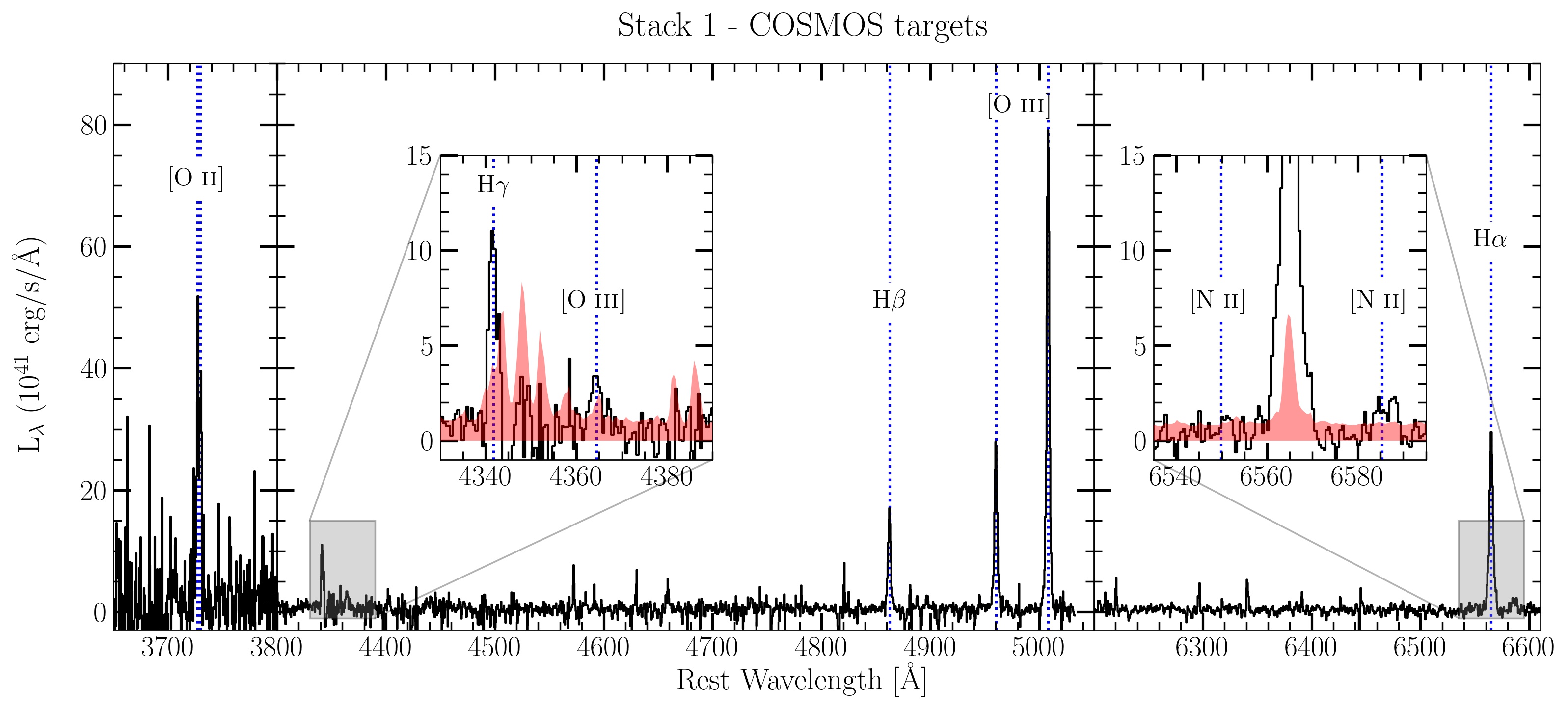}
    \includegraphics[width=17cm]{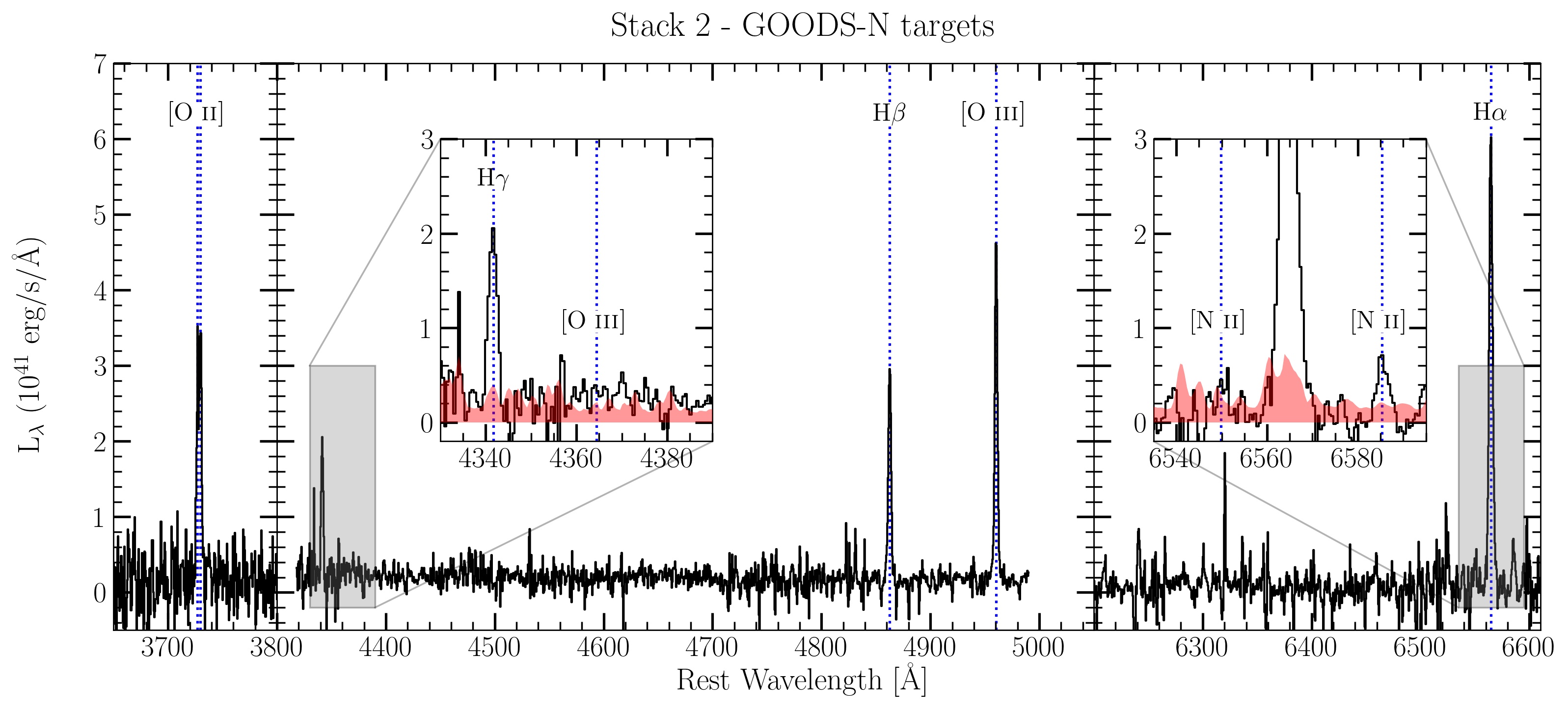}
    \includegraphics[width=17cm]{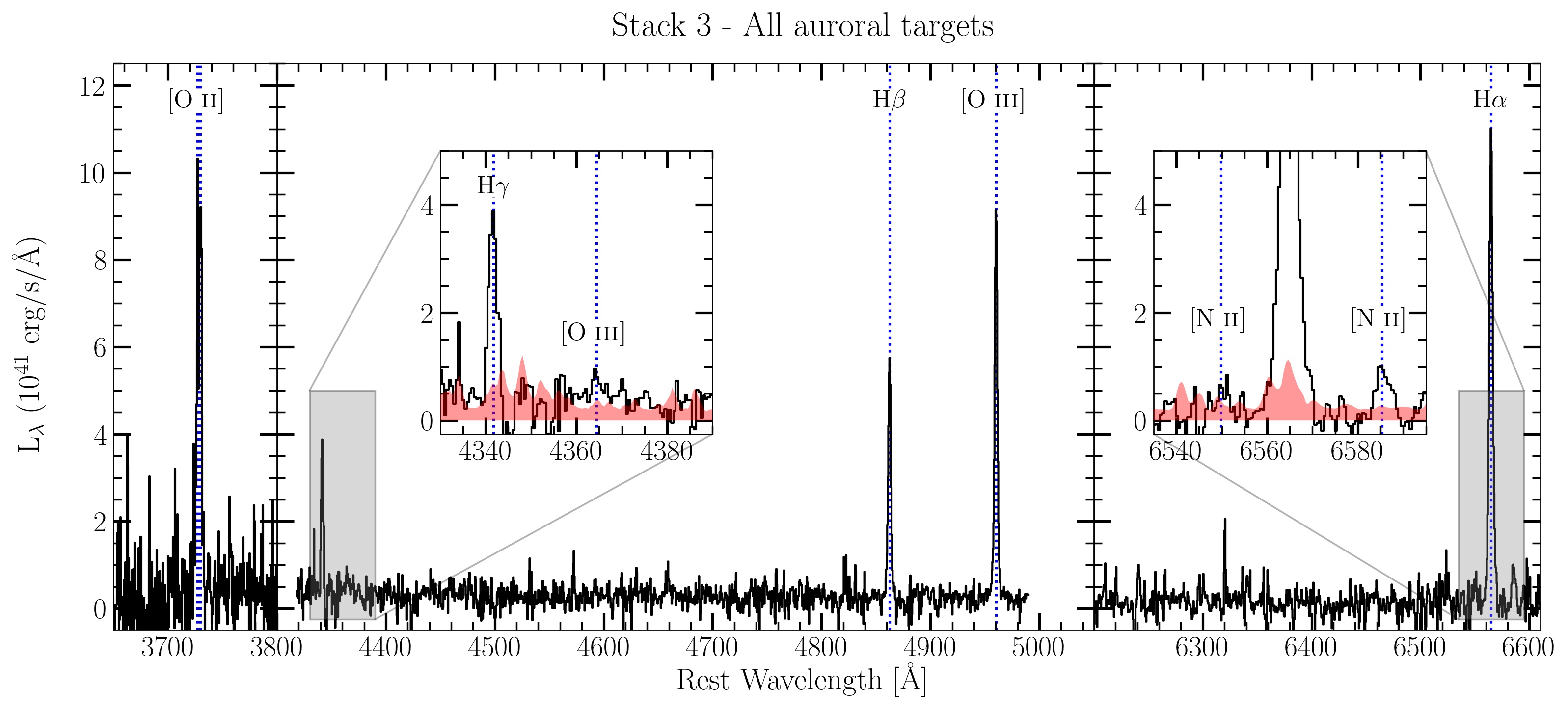}
    \caption{Composite spectra of the \oiii\ auroral targets. The black curve shows the luminosity density, while the error spectrum is shown in red in the inset axes. Each of the prominent emission lines is labeled and marked with a blue dotted line.}
    \label{fig:stacks}
\end{figure*}

In order to create the composite spectra, we first used the H$\alpha$/H$\beta$ Balmer decrement assuming Case B recombination \citep{1989agna.book.....O}, an intrinsic ratio of 2.86, and a \citet{1989ApJ...345..245C} extinction curve to correct for internal dust extinction. We then converted each spectrum from flux density ($F_{\lambda}$) to luminosity density ($L_{\lambda}$) by multiplying by $4\pi D_L^2(1+z)$ where $D_L$ is the luminosity distance. Subsequently, each spectrum was normalized by H$\alpha$ luminosity and shifted into the rest frame. Prior to stacking, each spectrum was interpolated and re-sampled to the same wavelength grid with 0.5 \AA\ spacing. The resulting stacked spectrum was then multiplied by the median H$\alpha$ luminosity of the component spectra in order to obtain units of luminosity density. We then averaged together the SED models of each component spectrum after normalizing by H$\alpha$ luminosity, and the resulting composite SED curve was used to model the continuum of the stacks during the line-fitting procedure. We report line luminosity measurements for each of the \oiii\ auroral stacks in Table \ref{tab:stacked_flux}. Additionally, we show the stacked spectra in Figure \ref{fig:stacks} with emission lines of interest labeled.

\begin{figure}[htb!]
    \centering
    \includegraphics[width=8cm]{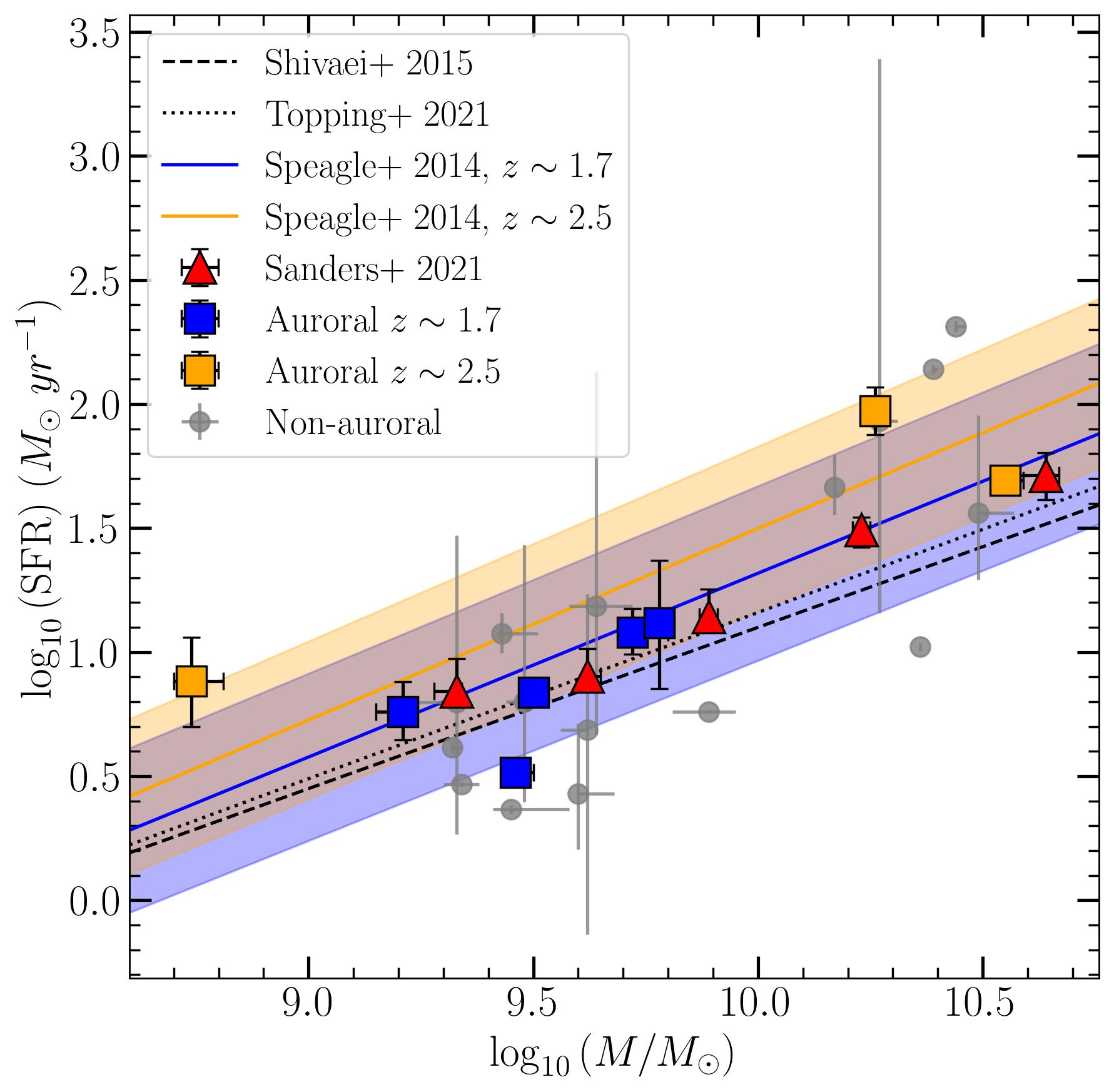}
    \caption{SFR vs. stellar mass. The colored squares indicate the auroral-line targets included in this study, with blue and orange corresponding to $z\sim 1.7$ and $z\sim 2.5$, respectively. The gray points indicate the non-auroral targets included in this study. The dashed black line shows a linear fit from \citet{2015ApJ...815...98S}, while the solid blue and orange lines show the $z\sim 1.7$ and $z\sim 2.5$ SFR vs. stellar mass relations respectively from \citet{2014ApJS..214...15S}. The 1-$\sigma $ scatter for each main sequence line is shaded in its respective color. The red triangles represent spectral stacks of $z\sim 2.3$ galaxies from \citet{2021ApJ...914...19S}.}
    \label{fig:sfr_vs_mass}
\end{figure}

\section{Results and Determination of Physical Quantities} \label{sec:results}

\begin{deluxetable*}{l|cccccccc}[t!]
    \tabletypesize{\scriptsize}
    \caption{Catalog of observed emission-line flux and physical properties for individual \oiii\ auroral targets.}
    \label{tab:all_flux}
    \tablehead{ \multicolumn{9}{c}{F$_{obs}(\lambda)(10^{-17}\ {\rm erg}\ {\rm s}^{-1}\ {\rm cm}^{-2}$)} \\
    \hline
    \multirow{2}{*}{Line} & \multicolumn{3}{c|}{ID (COSMOS)} & \multicolumn{5}{c}{ID (GOODS-N)}\\
    \cline{2-4} \cline{5-9}
     & \colhead{18812} & \colhead{19439} & \colhead{19753} \vline & \colhead{6699} & \colhead{8013} & \colhead{8240} & \colhead{14595} & \colhead{18462}
    }
    \startdata
        \oii\ $\lambda 3726$ & $<2.66$ & $2.78\pm0.71$ & $6.37\pm0.49$ & $2.18\pm0.26$ & $3.88\pm0.48$ & $1.92\pm0.42$ & $1.29\pm0.24$ & $1.74\pm0.25$ \\
        \oii\ $\lambda 3730$ & $0.50\pm0.23$ & $2.33\pm0.35$ & $8.21\pm0.55$ & $2.58\pm0.23$ & $4.30\pm0.66$ & $2.03\pm0.29$ & $1.38\pm0.21$ & $1.82\pm0.32$ \\
        \neiii\ $\lambda 3870$ & -- & $<1.52$ & $<1.19$ & $<0.60$ & $0.55\pm0.17$ & $<1.11$ & $<0.57$ & $<0.65$ \\
        H$\delta\ \lambda 4103$ & -- & -- & -- & $0.58\pm0.17$ & $1.22\pm0.30$ & $0.57\pm0.19$ & $<0.93$ & $<1.23$ \\
        H$\gamma\ \lambda 4342$ & $0.58\pm0.09$ & $1.12\pm0.14$ & $2.45\pm0.40$ & $1.83\pm0.21$ & $2.10\pm0.17$ & $1.59\pm0.14$ & $1.08\pm0.10$ & $1.35\pm0.20$ \\
        \feii\ $\lambda 4360^*$ & $<0.34$ & $0.18\pm0.08$ & $<0.31$ & $0.41\pm0.11$ & $<0.78$ & $<0.38$ & $<0.24$ & $<0.62$ \\
        \oiii\ $\lambda 4364^*$ & $0.34\pm0.09$ & $<0.28$ & $<0.27$ & $<0.51$ & $<0.99$ & $<0.29$ & $0.20\pm0.06$ & $<0.38$ \\
        \oiii\ $\lambda 4364$ & $0.34\pm0.11$ & $<0.28$ & $<0.27$ & $<0.51$ & $<0.99$ & $0.16\pm0.08$ & $<0.25$ & $<0.38$ \\
        H$\beta\ \lambda 4863$ & $1.10\pm0.15$ & $2.30\pm0.08$ & $5.88\pm0.10$ & $3.21\pm0.10$ & $4.77\pm0.13$ & $3.23\pm0.22$ & $2.24\pm0.10$ & $2.57\pm0.14$ \\
        \oiii\ $\lambda 4960$ & $2.49\pm0.07$ & $3.95\pm0.10$ & $7.47\pm0.22$ & $5.29\pm0.14$ & $5.71\pm0.19$ & $4.71\pm0.08$ & $3.79\pm0.08$ & $3.71\pm0.10$ \\
        \oiii\ $\lambda 5008$ & $6.99\pm0.11$ & $11.52\pm0.15$ & $21.49\pm0.12$ & $15.76\pm0.41$ & $15.61\pm0.10$ & $14.03\pm0.22$ & $11.11\pm0.12$ & $10.30\pm0.11$ \\
        \oi\ $\lambda 6302$ & $<0.40$ & $<0.48$ & $0.47\pm0.09$ & $0.26\pm0.11$ & $<0.40$ & $<1.35$ & $<0.35$ & $<0.34$ \\
        \oi\ $\lambda 6365$ & $<0.40$ & $<0.63$ & $0.15\pm0.03$ & $0.08\pm0.04$ & $<0.37$ & $<1.96$ & $<0.32$ & $<0.40$ \\
        \nii\ $\lambda 6550$ & $<0.37$ & $0.42\pm0.12$ & $1.01\pm0.21$ & $<1.85$ & $<0.79$ & $<0.76$ & $<0.61$ & $<0.80$ \\
        H$\alpha\ \lambda 6565$ & $3.28\pm0.14$ & $10.89\pm0.30$ & $20.21\pm0.21$ & $11.58\pm0.36$ & $12.51\pm0.79$ & $12.22\pm1.35$ & $7.07\pm0.17$ & $5.41\pm0.24$ \\
        \nii\ $\lambda 6585$ & $<0.33$ & $1.27\pm0.23$ & $3.03\pm0.26$ & $1.55\pm0.29$ & $1.24\pm0.28$ & $1.13\pm0.35$ & $<0.54$ & $0.55\pm0.23$ \\
        \sii\ $\lambda 6716$ & -- & $<0.61$ & $1.85\pm0.25$ & $<2.56$ & $1.90\pm0.68$ & $<3.12$ & $0.56\pm0.21$ & -- \\
        \sii\ $\lambda 6731$ & -- & $0.90\pm0.25$ & $1.66\pm0.26$ & $0.76\pm0.22$ & $<1.15$ & $<17.44$ & $<1.22$ & -- \\
        \hline
        E(B-V)$_{gas}$ & $0.04^{+0.16}_{-0.13}$ & $0.51^{+0.05}_{-0.05}$ & $0.19^{+0.02}_{-0.02}$ & $0.23^{+0.04}_{-0.04}$ & $0.00$ & $0.29^{+0.12}_{-0.14}$ & $0.10^{+0.05}_{-0.05}$ & $0.00$ \\
        $T_e$(O$^{2+}$) ($10^4$ K) & $2.11^{+0.29}_{-0.39}$ & $<1.42$ & $<1.13$ & $<2.41$ & $<2.29$ & $1.30^{+0.24}_{-0.25}$ & $<1.86$ & $<1.66$ \\
        $T_e$(O$^+$) ($\times 10^4$ K) & $1.77^{+0.26}_{-0.25}$ & $<1.30$ & $<1.09$ & $<1.99$ & $<1.90$ & $1.20^{+0.20}_{-0.20}$ & $<1.61$ & $<1.46$ \\
        $n_e\ (10^2\ \rm cm^{-3}$) & -- & $10.70^{+12.55}_{-6.84}$ & $1.61^{+1.32}_{-0.98}$ & $4.20^{+2.20}_{-2.79}$ & $5.26^{+4.77}_{-3.34}$ & $5.06^{+4.97}_{-3.26}$ & $5.38^{+6.53}_{-3.12}$ & $5.57^{+6.11}_{-3.41}$ \\
        $12+\log{({\rm O}^+/{\rm H})}$ & -- & $>7.78$ & $>7.89$ & $>6.89$ & $>6.85$ & $7.49^{+0.32}_{-0.24}$ & $>6.99$ & $>6.98$ \\
        $12+\log{({\rm O}^{2+}/{\rm H})}$ & $7.54^{+0.19}_{-0.13}$ & $>7.77$ & $>7.94$ & $>7.27$ & $>7.18$ & $7.82^{+0.28}_{-0.19}$ & $>7.50$ & $>7.56$ \\
        $12+\log{({\rm O}/{\rm H})}$ & -- & $>8.08$ & $>8.22$ & $>7.42$ & $>7.35$ & $8.02^{+0.24}_{-0.17}$ & $>7.62$ & $>7.66$ \\
        $12+\log{({\rm N}^+/{\rm H})}$ & $5.83^{+0.30}_{-0.47}$ & $>6.57$ & $>6.85$ & $>6.28$ & $>6.18$ & $6.53^{+0.24}_{-0.21}$ & $>5.85$ & $>6.39$ \\
        $12+\log{({\rm N}/{\rm H})}$ & $6.71^{+0.52}_{-0.61}$ & $>6.86$ & $>7.18$ & $>6.82$ & $>6.67$ & $7.05^{+0.29}_{-0.26}$ & $>6.49$ & $>7.07$ \\
        $\log{({\rm N}/{\rm O})}$ & -- & -- & -- & -- & -- & $-0.99^{+0.22}_{-0.23}$ & -- & -- \\
    \enddata
    \tablenotetext{*}{These line fluxes for \feii $\lambda 4360$ and \oiii $\lambda 4364$ are determined by fitting a double Gaussian simultaneously to both features, whereas the \oiii $\lambda 4364$ flux without an asterisk is determined by fitting a single Gaussian profile.}
    \tablenotetext{**}{  Since the \oii\ doublet was affected by sky lines in this target, we estimate the $\rm O^{2+}/H$ abundance assuming a density of 100 $\rm cm^{-3}$.}
    \tablecomments{Non-detections are reported as 2$\sigma$ upper limits, and Balmer line ratios that yield negative values of E(B-V) are set to 0.00. Oxygen abundances are determined from the single-Gaussian fit to \oiii $\lambda 4364$.}
\end{deluxetable*}

\begin{deluxetable}{l|ccc}
    \tabletypesize{\scriptsize}
    \caption{Catalog of observed emission-line luminosities and physical properties for stacked spectra.}
    \label{tab:stacked_flux}
    \tablehead{ \multicolumn{4}{c}{L$_{obs}(\lambda)(10^{41}\ erg\ s^{-1}$)} \\
    \hline
    \multirow{3}{*}{Line} & \multicolumn{3}{c}{Stack ID}\\
    \cline{2-4}
     & \colhead{S1} & \colhead{S2} & \colhead{S3}\\
     & \colhead{COSMOS} & \colhead{GOODS-N} & \colhead{Full sample}
    }
    \startdata
        \oii\ $\lambda 3726$ & $88.13\pm28.17$ & $6.78\pm1.17$ & $19.58\pm5.00$ \\
        \oii\ $\lambda 3730$ & $89.12\pm24.34$ & $8.15\pm1.11$ & $21.86\pm3.36$ \\
        \neiii\ $\lambda 3870$ & -- & $<1.36$ & -- \\
        H$\delta\ \lambda 4103$ & -- & $<1.60$ & -- \\
        H$\gamma\ \lambda 4342$ & $27.21\pm4.52$ & $4.46\pm0.52$ & $9.23\pm0.90$ \\
        \feii\ $\lambda 4360^*$ & $<12.91$ & $<1.15$ & $1.87\pm0.84$ \\
        \oiii\ $\lambda 4364^*$ & $7.56\pm3.21$ & $<0.76$ & $1.40\pm0.46$ \\
        \oiii\ $\lambda 4364$ & $7.57\pm2.71$ & $<0.50$ & $1.39\pm0.40$ \\
        H$\beta\ \lambda 4863$ & $49.78\pm4.91$ & $9.16\pm0.50$ & $17.38\pm0.94$ \\
        \oiii\ $\lambda 4960$ & $76.66\pm8.72$ & $13.19\pm0.90$ & $26.04\pm1.33$ \\
        \oiii\ $\lambda 5008$ & $255.96\pm24.15$ & -- & -- \\
        \oi\ $\lambda 6302$ & $<5.08$ & $<0.74$ & $<0.81$ \\
        \oi\ $\lambda 6365$ & $<5.44$ & $<0.25$ & $<0.27$ \\
        \nii\ $\lambda 6550$ & $3.21\pm0.74$ & $0.62\pm0.10$ & $1.08\pm0.16$ \\
        H$\alpha\ \lambda 6565$ & $110.87\pm6.29$ & $24.01\pm1.25$ & $43.23\pm1.47$ \\
        \nii\ $\lambda 6585$ & $9.63\pm2.23$ & $1.85\pm0.29$ & $3.25\pm0.47$ \\
        \sii\ $\lambda 6716$ & -- & -- & -- \\
        \sii\ $\lambda 6731$ & -- & -- & -- \\
        \hline
        $\log{(M_*/M_{\odot,avg})}$ & $9.85\pm 0.04$ & $9.51\pm 0.06$ & $9.64\pm 0.04$\\
        $T_e$(O$^{2+}$) ($10^4$ K) & $1.96^{+0.33}_{-0.37}$ & $<1.27$ & $1.44^{+0.19}_{-0.19}$ \\
        $T_e$(O$^{+}$) ($10^4$ K) & $1.67^{+0.26}_{-0.26}$ & $<1.19$ & $1.30^{+0.15}_{-0.15}$ \\
        $n_e$ ($10^2$ cm$^{-3}$) & $8.68^{+14.23}_{-6.29}$ & $3.14^{+3.55}_{-2.14}$ & $4.65^{+5.99}_{-3.17}$ \\
        $12+\log{(\text{O}^+/\text{H})}$ & $7.41^{+0.26}_{-0.22}$ & $>7.49$ & $7.53^{+0.22}_{-0.18}$ \\
        $12+\log{(\text{O}^{2+}/\text{H})}$ & $7.43^{+0.21}_{-0.16}$ & $>7.85$ & $7.72^{+0.17}_{-0.15}$ \\
        $12+\log{(\text{O}/\text{H})}$ & $7.75^{+0.20}_{-0.15}$ & $>8.01$ & $7.96^{+0.15}_{-0.12}$ \\
        $12+\log{(\text{N}^+/\text{H})}$ & $6.13^{+0.16}_{-0.17}$ & $>6.44$ & $6.31^{+0.15}_{-0.12}$ \\
        $12+\log{(\text{N}/\text{H})}$ & $6.46^{+0.18}_{-0.18}$ & $>6.96$ & $6.73^{+0.12}_{-0.11}$ \\
        $\log{(\text{N}/\text{O})}$ & $-1.30^{+0.17}_{-0.19}$ & -- & $-1.22^{+0.11}_{-0.12}$ \\
    \enddata
    \tablenotetext{*}{These line luminosities for \feii $\lambda 4360$ and \oiii $\lambda 4364$ are determined by fitting a double Gaussian simultaneously to both features, whereas the \oiii $\lambda 4364$ luminosity without an asterisk is determined by fitting a single Gaussian profile. Oxygen abundances are determined from the single-Gaussian fit to \oiii $\lambda 4364$.}
    \tablecomments{Non-detections are reported as 2$\sigma$ upper limits.}
\end{deluxetable}

\subsection{SFR vs. stellar mass}

In Figure \ref{fig:sfr_vs_mass}, we plot the SFRs vs. stellar masses for our galaxy sample and compare them to typical values from the literature. The eight auroral targets shown in Table \ref{tab:galaxies} have H$\alpha$ and H$\beta$ coverage, and we include them in the figure. Only 16 of the 29 filler targets, however, have simultaneous H$\alpha$ and H$\beta$ detections, and these are included as gray points in Figure \ref{fig:sfr_vs_mass}. We calculate SFRs from dust-corrected H$\alpha$ luminosities using a conversion factor of $3.236\times 10^{-42}$ M$_{\odot}$ yr$^{-1}$ erg$^{-1}$ s based on models of stellar populations with sub-solar metallicity consistent with what we assume for SED fitting \citep{2018ApJ...869...92R}. The galaxies in this sample agree well with the star-forming main sequence defined by larger populations of galaxies at similar redshifts. \citet{2015ApJ...815...98S} performed a linear fit to a sample of 185 galaxies in the range $1.37 \leq z \leq 2.61$, and \citet{2021MNRAS.506.1237T} fit a sample of 285 galaxies in the range $1.37 \leq z \leq 1.70$. These two empirical fits are shown as dashed and dotted black lines respectively, and they agree with the five GOODS-N galaxies at $z\sim 1.7$ with the exceptions of GOODS-N-18462 and GOODS-N-14595, which are offset in log(SFR) by roughly -0.3 dex and +0.6 dex respectively. We additionally plot two main sequence relations defined by \citet{2014ApJS..214...15S} at the median redshifts of our COSMOS and GOODS-N samples shown in orange and blue, respectively. As an additional point of reference, we overplot a sample of 280 $z\sim 2.3$ galaxies distributed among five stacks by \citet{2021ApJ...914...19S}, and we re-scale the stacks to use the same $H\alpha$ to SFR conversion factor we utilize in this study. For both the $z \sim 1.7$ and $z \sim 2.5$ galaxies in this study, we find agreement within 1$\sigma$ relative to the respective main sequence fits with the exception of GOODS-N-18462, which falls slightly below the $z \sim 1.7$ relation. Overall, this sample of galaxies is relatively representative in terms of SFR at fixed stellar mass based on larger samples in the same redshift range.

\subsection{Abundance Determinations}

Throughout this paper, we refer to ``direct" oxygen abundances as those derived from determining the ion emissivities based on electron temperature and density measurements as opposed to ``indirect" methods, which use the ratios of strong nebular emission lines empirically calibrated to local direct measurements or photoionization models. The direct method of oxygen abundance approximates a galaxy as a single \hii region and thus characterizes electron temperatures and densities based on globally-integrated spectra. In this \hii region approximation, it is conventional to further define two temperatures: the temperature associated with the high-ionization state and the low-ionization state of the ions of interest (e.g., $T_e(\text{O}^{2+})$ and $T_e(\text{O}^+)$, respectively). Since the energy required to ionize neutral oxygen is similar to that of hydrogen, we expect that nearly all of the oxygen within \hii regions is ionized, either in the singly or doubly-ionized state, with negligible amounts in higher ionization or neutral states. Therefore, in order to determine the oxygen abundance directly, we sum the abundances of oxygen in its two most prevalent ionization states:

\begin{equation}
    \rm \frac{O}{H}\approx\frac{O^+}{H^+}+\frac{O^{2+}}{H^+}
\end{equation}

Determining the abundances of each ionization species requires knowledge of the electron temperatures and densities associated with the respective ionization zones. In the O$^{2+}$ zone, this can be achieved through measurements of the \oiii $\lambda\lambda 4960,5008$ and the \oiii $\lambda 4364$ lines, where the \oiii $\lambda 4364$ transition originates from a different upper energy level than the \oiii $\lambda 4960,5008$ transitions. In turn, measuring the ratios of these lines allows one to determine the electron temperature associated with the O$^{2+}$ zone. The same can be achieved for the O$^+$ ion, using the \oii $\lambda\lambda 3727,3730$ and \oii $\lambda\lambda 7321,7332$ lines which arise from different respective upper energy levels. One of the major challenges in determining direct oxygen abundances, however, is that the auroral lines produced by transitions from the upper energy levels are often intrinsically very faint, and their detection becomes the main limiting factor in obtaining direct abundance estimates. Ideally, one would measure the nebular and the corresponding faint auroral emission lines originating from the O$^+$ and O$^{2+}$ ions to directly determine the electron temperature for both ionization zones. However, since our sample of \oiii\ auroral targets does not have coverage of the auroral \oii\ lines, we employ the following theoretical relation presented by \citet{1986MNRAS.223..811C} to infer $T_e(\text{O}^+)$:

\begin{equation}\label{eq:campbell}
    T_e(\text{O}^+) = 0.7\times T_e(\text{O}^{2+}) + 3000\thinspace \text{K}
\end{equation}

When converting a $T_e$(O$^{+}$) measurement to $T_e$(O$^{2+}$), observations suggest an intrinsic scatter of approximately 1300 K in the \citet{1986MNRAS.223..811C} relation \citep{2020ApJ...893...96B,2021ApJ...915...21R}. Since we are instead converting $T_e$(O$^{2+}$) to $T_e$(O$^{+}$) using equation \ref{eq:campbell}, we adopt an intrinsic scatter of $0.7\times 1300\ \text{K}=910\ \text{K}$, and add this in quadrature when determining the uncertainty on $T_e$(O$^{+}$).

Electron temperatures, densities, and ionic abundances were determined using the \texttt{PyNeb} package \citep{2015A&A...573A..42L}. In order to compute the electron density $n_e$, we used the \texttt{getCrossTemDen()} method to simultaneously solve for $T_e$(O$^{2+}$) and $n_e$, taking the \oii $\lambda 3727$/\oii $\lambda 3730$ ratio to be the density-sensitive tracer. With the output values of $T_e$ and $n_e$ from \texttt{getCrossTemDen()}, we computed the ionic abundances using the \texttt{getIonAbundance()} method, using the ratios of \oiii $\lambda 4959$ and \oii $\lambda 3727,3730$ relative to H$\beta$ to compute the $ \rm O^{2+}/H^+$ and the $\rm O^+/H^+$ abundances, respectively.

We additionally calculated the nitrogen abundance in our galaxy sample since we have coverage of the \nii $\lambda 6585$ line in all targets. The nitrogen abundance is ideally determined based on emission lines arising from the N$^+$ and N$^{2+}$ ions. However, we do not have coverage and/or detection of the necessary \niii\ emission lines, so we make the following approximation:

\[
\rm \frac{N}{H} \approx \frac{N^+}{H^+}\times ICF(N)
\]

\begin{figure*}
    \centering
    \includegraphics[width=18cm]{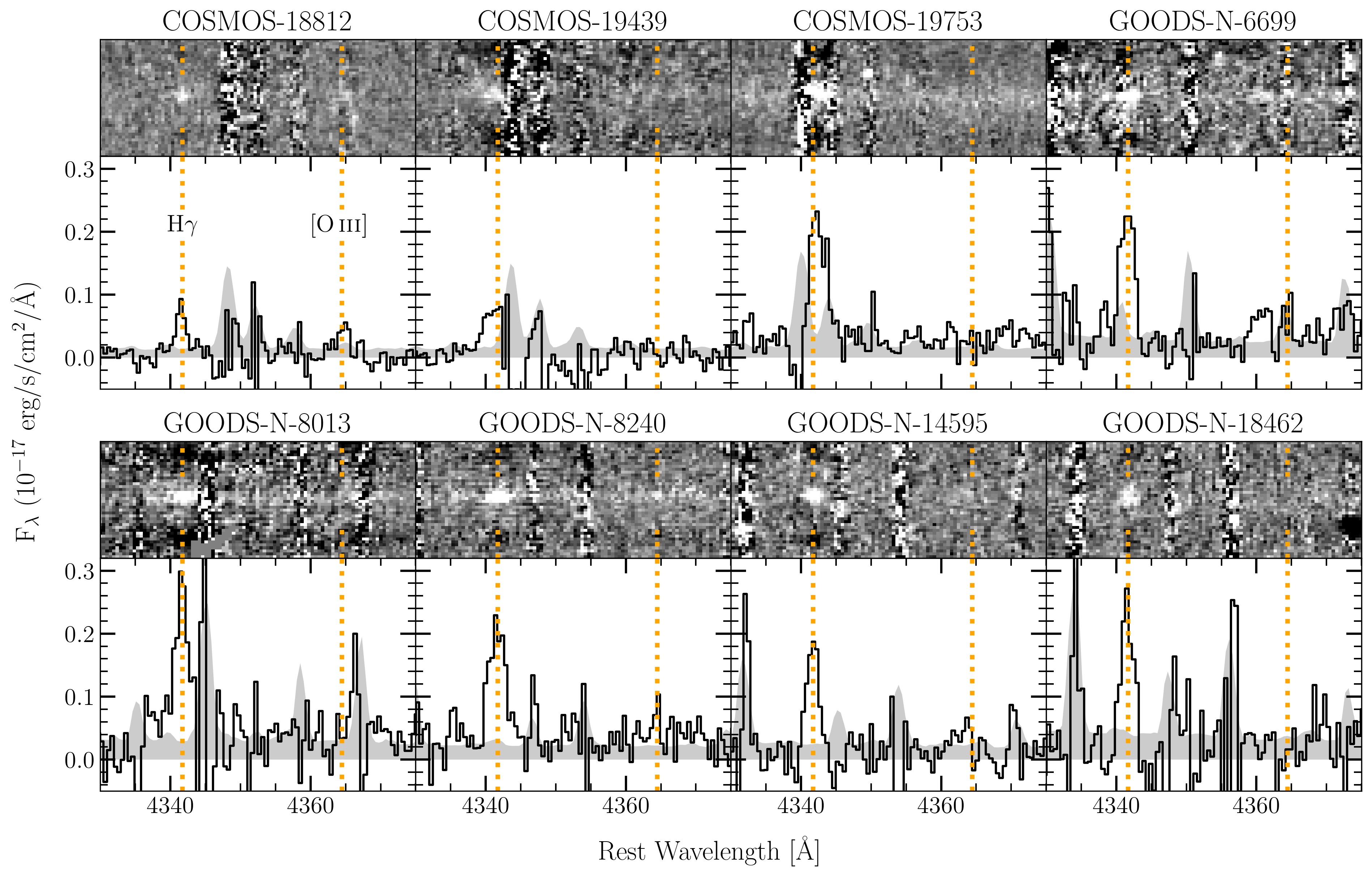}
     \caption{2D and 1D spectra of the eight \oiii\ auroral targets in this study. These spectral plots span from 4335 \AA\ to 4370 \AA, covering the H$\gamma$ and \oiii $\lambda 4364$ emission lines. The solid black line shows the flux density, and the shaded gray spectrum is the error on the flux density. We detect \oiii $\lambda 4364$ (labeled with an orange dotted line) in COSMOS-18812 and GOODS-N-8240; however, we report emission blueward of 4364 \AA\ in the GOODS-N 6699 and 14595 spectra.}
    \label{fig:auroral_lines}
\end{figure*}

where ICF(N) is the ionization correction factor accounting for higher ionization states, and it is defined as $\rm ICF(N)=N/N^+$. We approximate this ratio as $\rm N/N^+\approx O/O^+$ since oxygen and nitrogen have similar ionization energies \citep{1967ApJ...150..825P}.

One can directly measure the temperature within the $\rm N^+$ zone by measuring the auroral-to-nebular line ratio \nii $\lambda 5756$/\nii $\lambda 6585$, analogous to the \oiii $\lambda 4364$/\oiii $\lambda 5008$ ratio for the $\rm O^{2+}$ zone. However, we do not have coverage of the \nii $\lambda 5756$ line in our spectra, so we make the approximation that $T_e( {\rm N}^+)\approx T_e({\rm O}^+)$ since both ions should occupy the low-ionization zones. We also use the same electron density as that determined during the calculation of the oxygen abundance. We then used \texttt{getIonAbundance()} employing the \nii $\lambda 6585$/H$\beta$ ratio as the input to calculate the $\rm N^+/H^+$ abundance. The derived constraints on density, temperature, and ionic and total abundances are reported in Table \ref{tab:all_flux} for the individual targets and Table \ref{tab:stacked_flux} for the composites.

\subsection{Auroral-line Detections}

The \oiii $\lambda 4364$ line was detected at greater than 2$\sigma$ significance in only two targets: COSMOS-18812 and GOODS-N-8240. For COSMOS-18812, the \oii $\lambda\lambda 3727,3730$ line doublet fell onto a pair of sky lines, so it was not possible to constrain the electron density or the O$^+$ abundance. We tentatively detect \oiii $\lambda4364$ emission in GOODS-N-8240 since the Gaussian fit to the line profile places the significance of this detection at the $>$2$\sigma$ level. Upon visual inspection, we report the presence of emission lying a few angstroms blueward of the \oiii $\lambda 4364$ line in two of the targets: GOODS-N-6699 and GOODS-N-14595. For GOODS-N-6699, there is an emission feature detected at $>$3$\sigma$ significance when centering a Gaussian profile at 4360 \AA. It is uncertain whether this emission is associated with the \oiii $\lambda 4364$ line though it appears at a similar wavelength. The emission-line feature adjacent to \oiii $\lambda 4364$ in GOODS-N-14595 appears to be more closely centered on the expected central wavelength for the auroral oxygen line, with a significance of just 0.38$\sigma$ when centering on 4360 \AA. The spectra of these objects in this wavelength region can be seen in Figure \ref{fig:auroral_lines}.

\citet{2017MNRAS.465.1384C} find a similar feature in a sample of spectral stacks and attribute it to a forbidden transition of singly-ionized iron. They also suggest that the strength of this 4360 \AA\ contamination increases with increasing galaxy metallicity. However, we do not expect this contaminating feature to have a significant impact on our spectra given that strong-line metallicity indicators of our target sample suggest our targets are half solar metallicity or less. For completeness, we report \oiii $\lambda 4364$ measurements of the stacked spectra in two ways: fitting the emission around 4364 \AA\ as a single Gaussian and fitting the feature at 4360 \AA\ separately from \oiii $\lambda 4364$. As a shorthand, we refer to the feature at 4360 \AA\ as \feii $\lambda 4360$, and the simultaneous \oiii\ and \feii\ line fits are denoted by an asterisk (*) in both Tables \ref{tab:all_flux} and \ref{tab:stacked_flux}.

In Table \ref{tab:all_flux}, we see that in most cases, choosing a single vs. a double fit does little to change the \oiii $\lambda 4364$ flux. The exceptions to this are GOODS-N-8240 and GOODS-N-14595 where, in the former, the single fit yields a higher S/N ratio and, in the latter, the double fit yields a better S/N ratio. Since it is unclear to what extent the emission in GOODS-N-14595 can be attributed to \oiii $\lambda 4364$, we report the determination of physical quantities from the single-Gaussian fit to \oiii $\lambda 4364$.

We perform this same exercise on the stacked spectra and report the results in Table \ref{tab:stacked_flux}, with the simultaneous double Gaussian fits marked by asterisks. Since the emission at 4360 \AA\ is only seen in GOODS-N targets, we check to see if the fitting technique has an effect on the measured line luminosities in stacks 2 and 3, finding that there is no significant effect on either stack. We do see an effect on stack 1 in that the single fit has a higher S/N ratio. Thus, we use the single Gaussian fit to \oiii $\lambda 4364$ in the stacks to determine physical conditions.

For the individual \oiii\ auroral targets, only GOODS-N-8240 has a well-constrained oxygen abundance of $\rm 12+\log(O/H)=8.02^{+0.24}_{-0.17}$, corresponding to $\sim$21\% of the solar oxygen abundance. For the composite spectra, we find that the $z\sim 2.5$ COSMOS stack has an oxygen abundance of $\sim 11\%$ the solar value, while the $z\sim 1.7$ GOODS-N stack has an oxygen abundance greater than $\sim 21 \%$ the solar value. For nitrogen, the abundances relative to the solar value are $\sim 4\%$ and $\gtrsim 13\%$ for stacks 1 $(z\sim 2.5)$ and 2 $(z\sim 1.7)$.

\begin{figure*}
    \centering
    \includegraphics[width=16cm]{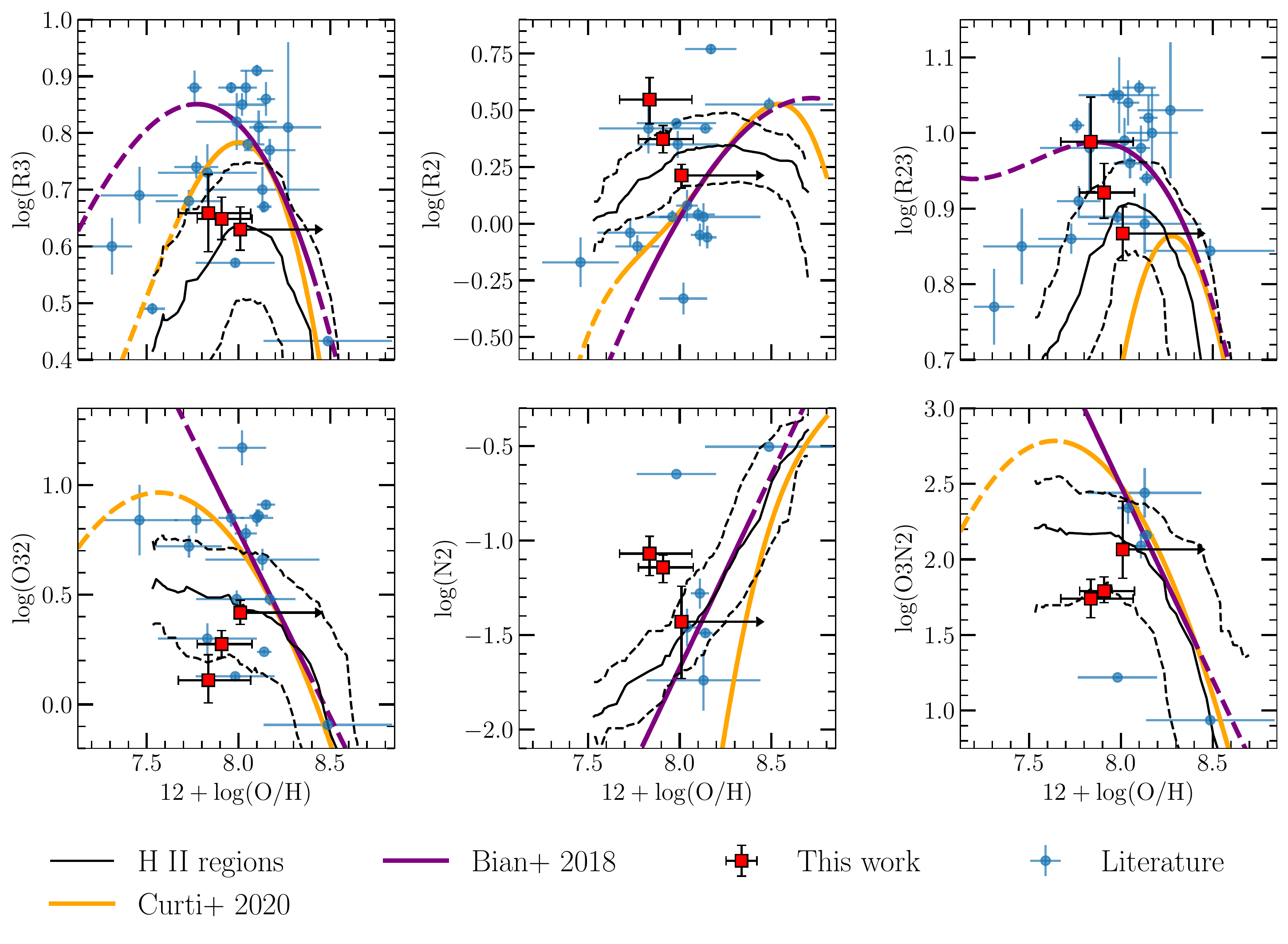}
    \caption{Comparison of the direct oxygen abundance measurements of our stacks with strong-line indicators from \citet{2020MNRAS.491..944C} in orange and \citet{2018ApJ...859..175B} in purple. Strong-line indicators are plotted in solid lines over their quoted metallicity ranges, and extrapolations are shown by dashed lines. We also display $z>1$ galaxies from the literature with direct oxygen abundance measurements as blue points \citep[see][]{2020MNRAS.491.1427S,2023ApJ...943...75S}. Finally, we show the median relation of local \hii regions \citep[see][]{2017ApJ...850..136S} as a solid black curve, with the 1$\sigma$ spread illustrated as black dotted curves.}
    \label{fig:line_ratios}
\end{figure*}

\section{Discussion}\label{sec:discussion}

We now turn to an analysis of the emerging trends in relations between strong-line ratios and direct oxygen abundance at high redshift enabled by our deep MOSFIRE observations. In addition, the new analysis of the \oi $\lambda 6302$/H$\alpha$ ratio in 21 galaxies beyond the local universe, while not representing a complete statistical sample, hints at the properties of the ionized gas and the stellar populations in high-redshift galaxies.

\subsection{Indirect Metallicity Indicators}

Analyzing the accuracy of indirect metallicity indicators out to high redshifts is important for our understanding of the chemical evolution of galaxies across cosmic time. With our oxygen abundance measurements of the stacked spectra of auroral targets, we have constraints for the average metallicities of the galaxies considered in this study. We compare these stacks alongside measurements from the literature to strong-line metallicity indicators in order to understand how the accuracy of these indicators may shift with cosmic time.

In Figure \ref{fig:line_ratios}, we show six strong-line ratios vs. oxygen abundance for our three stacks, and we plot the metallicity relations from \citet{2020MNRAS.491..944C} determined from stacks of galaxy spectra in the local universe. The short-hand labels for the strong-line ratios are defined as follows: R3 = \oiii $\lambda 5008$/H$\beta$, R2 = \oii $\lambda\lambda 3727,3730$/H$\beta$, R23 = (\oiii $\lambda\lambda 5008, 4960$ + \oii $\lambda\lambda 3727, 3730$)/H$\beta$, O32 = \oiii $\lambda 5008$/\oii $\lambda\lambda 3727,3730$, N2 = \nii $\lambda 6585$/H$\alpha$, O3N2 = (\oiii $\lambda 5008$/H$\beta$)/(\nii $\lambda 6585$/H$\alpha$). We note that below an oxygen abundance of $12+\log{(\text{O}/\text{H})}\approx 8.1$, there are fewer individual $z\sim 0$ SDSS galaxies with $>$$10 \sigma$ \oiii $\lambda 4364$ detections, and the sample is biased toward higher specific SFR, representing a population more similar to our high-redshift sample than $z\sim 0$ galaxies (refer to the discussion in the appendix of \citet{2021ApJ...914...19S} for a detailed analysis of the low-metallicity \citet{2017MNRAS.465.1384C,2020MNRAS.491..944C} calibration sample).
Additionally, we show the relationships between strong-line ratios and metallicity determined by \citet{2018ApJ...859..175B} for local galaxies selected to have emission-line properties analogous to those of high-redshift galaxies.
Alongside these two line-ratio relations, we show the strong-line ratio vs. oxygen abundance for a large sample of \hii regions from the literature (compiled by \citet{2017ApJ...850..136S} with data from \citet{2016MNRAS.457.3678P}, \citet{2015ApJ...808...42C}, and \citet{2016MNRAS.458.1866T}) with a running median displayed as a solid black curve, and 1$\sigma$ intervals shown as dotted lines to visualize the spread of the distribution.

Upon visual inspection, the oxygen abundances of the stacks lie within the distribution of galaxies from the literature (compiled by \citet{2020MNRAS.491.1427S} with two additional galaxies from \citet{2023ApJ...943...75S}) shown as blue points. In the cases of log(R23) and log(N2), the \citet{2018ApJ...859..175B} curve serves as a better metallicity indicator to high-redshift galaxies compared to the \citet{2020MNRAS.491..944C} curves. In the cases of the log(R3), log(R2), log(O3O2), and log(O3N2) curves, the spread of galaxies is large compared to the differences between the \citet{2020MNRAS.491..944C} and \citet{2018ApJ...859..175B} curves, so it is difficult to determine if there is a preference for one over the other. In general, the stacks agree most consistently with the distribution of \hii regions, though the galaxies from the literature are offset from the \hii regions to higher log(R3) and log(R23) at fixed oxygen abundance.

With a small existing sample size, it is difficult to make definitive conclusions about the accuracy of these strong-line ratios, especially considering that the sample of galaxies is biased towards bright, high electron temperature targets. For two of the individual auroral-line targets where there were existing MOSDEF spectra (COSMOS-19439 and COSMOS-19752), we predicted the \oiii $\lambda 4364$ flux based on \oiii $\lambda 5008$ flux as well as $T_e$ predictions. For COSMOS-19439 and COSMOS-19753, we predicted auroral \oiii\ line fluxes in the ranges of $0.8-1.9 \times 10^{-18}$ and $2.1-5.0 \times 10^{-18}$ erg s$^{-1}$ cm$^{-2}$ respectively. Since the 2$\sigma$ upper limits are above our lowest line flux predictions, this suggests that the observations did not reach the required depth in the 10 combined hours of integration. This comparison demonstrates the limitations of 10-m-class ground-based observatories in this area of study and highlights the importance of JWST in building representative galaxy samples moving forward.

\subsection{Insights from [O\thinspace {\sc{i}}] emission}

\begin{figure*}
    \centering
    \includegraphics[width=18cm]{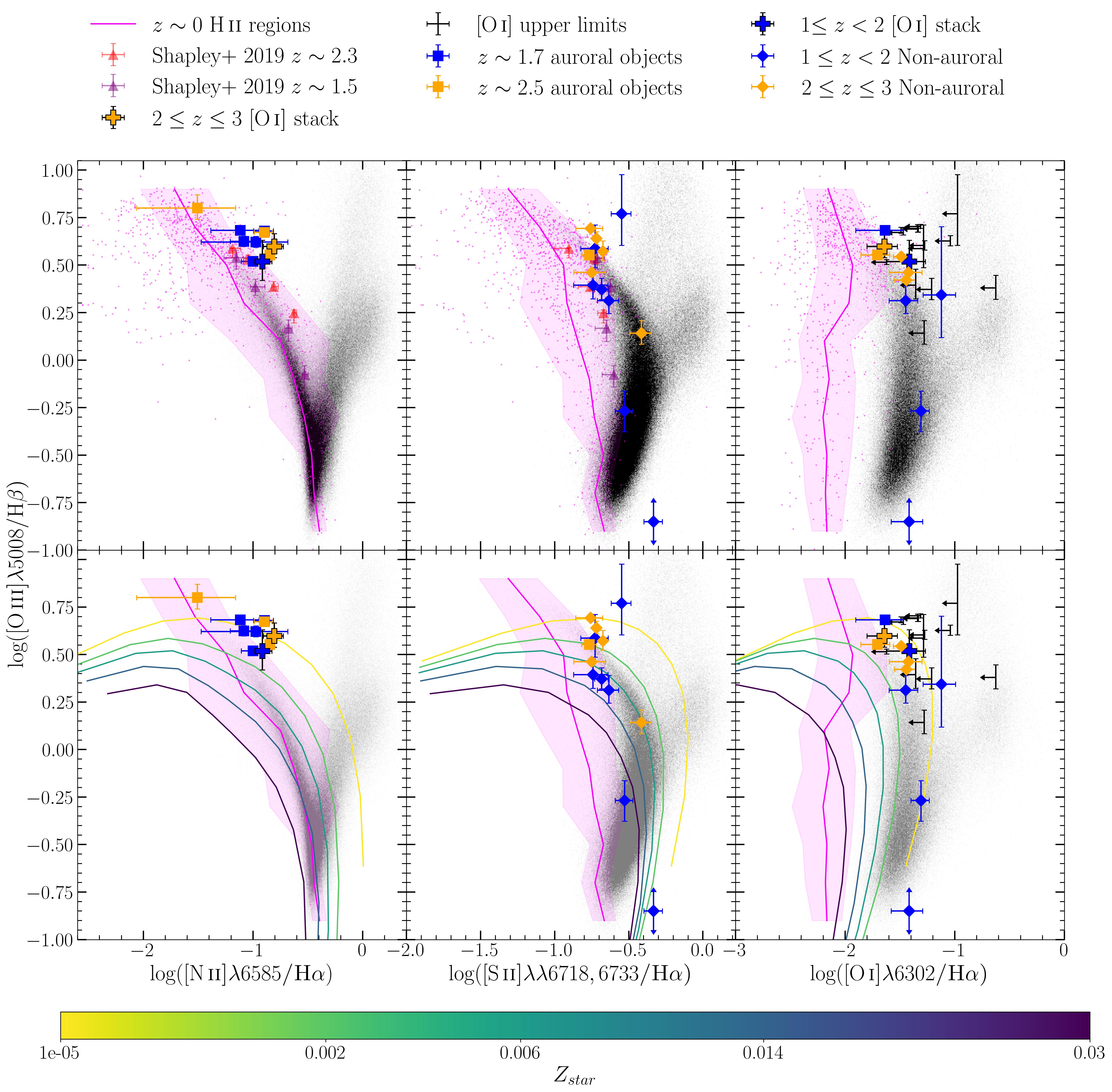}
    \caption{\nii, \sii, and \oi\ BPT diagrams showing where the auroral and \oi\ galaxy samples from this work lie in relation to SDSS galaxies in the local universe (grayscale, 2D histogram). We also compare the galaxies and stacks from this work to local \hii regions from the literature (compiled by \citet{2017ApJ...850..136S}) and display a running median. The \oiii\ auroral targets are shown by squares, while the non-\oiii-auroral targets are shown by diamonds. The $1\leq z < 2$ and $2\leq z \leq 3$ targets are displayed in blue and orange respectively. For comparison, stacks of $z\sim 1.5$ and $z\sim 2.3$ galaxies from \citet{2019ApJ...881L..35S} are shown as purple and red triangles respectively. Upper limits on \sii\ and \oi\ detections are shown in black. In the bottom three panels, the same BPT diagrams are shown with CLOUDY photoionization models overlaid. The curves are color-coded by stellar metallicity indicated in the colorbar.}
    \label{fig:bpt}
\end{figure*} 

We present the properties of our sample of galaxies in the \oiii $\lambda 5008$/H$\beta$ vs. \nii$\lambda 6585$/H$\alpha$, \sii$\lambda\lambda 6716, 6731$/H$\alpha$, and \oi $\lambda 6302$/H$\alpha$ BPT diagrams shown in Figure \ref{fig:bpt}. In all diagrams, local Sloan Digital Sky Survey \citep[SDSS;][]{2000AJ....120.1579Y,2009ApJS..182..543A} galaxies are shown as grayscale, two-dimensional histograms. A sample of local \hii regions from the literature \citep[see][]{2017ApJ...850..136S} is shown as a set of magenta points with an accompanying running median and a 1$\sigma$ shaded region. In the \nii\ BPT diagram, we plot the \oiii\ auroral targets as well as the \oi\ composite spectra. We see that the \oiii\ auroral sample (represented by squares) consists of high-excitation galaxies (log(\oiii$\lambda 5008$/H$\beta$) $\gtrsim 0.5$) and skews toward higher \oiii/H$\beta$ at fixed \nii/H$\alpha$ compared to the $z\sim 1.5$ and $z \sim 2.3$ samples from \citet{2019ApJ...881L..35S}. When compared to \hii regions in the literature, the auroral targets are $\sim$0.1 dex higher in log(\nii$\lambda 6585$/H$\alpha$) than the median locus of \hii regions at fixed log(\oiii/H$\beta$). The characteristics are similar for the $z\sim 2.5$ \oiii\ auroral galaxy in the \sii/H$\alpha$ diagram, where it is offset from the \hii regions and \citet{2019ApJ...881L..35S} samples at higher \sii/H$\alpha$ and \oiii/H$\beta$.

We additionally present a novel analysis of galaxies on the \oi\ BPT diagram at $z>1$. Including the filler targets, a total of nine galaxies yielded significant ($>$2$\sigma$) \oi $\lambda 6302$ detections, with two of the auroral \oiii\ targets (COSMOS-19753 and GOODS-N-6699) yielding detections. In order to understand the general characteristics of the galaxies in regard to the \oi\ BPT diagram, we constructed two composite spectra separated by redshift, choosing to include galaxies with coverage of the \oi\ lines with the exception of galaxies whose \oi\ feature fall on sky lines. These criteria result in two stacks with 13 galaxies in the $1\leq z < 2$ stack and 8 galaxies in the $2\leq z \leq 3$ stack. The line ratios associated with these stacks are plotted as ``plus" (+) symbols in Figure \ref{fig:bpt}.

We see that these stacks follow a similar trend of relatively high \oiii/H$\beta$ relative to the SDSS sample and high \oi/H$\alpha$ relative to the locus of \hii regions. There are several factors that can influence the \oi/H$\alpha$ ratio in a galaxy, including contributions of DIG, the presence of shocks, and hardness of the ionizing spectrum, the latter of which appears to be relevant in $z>1$ galaxies \citep{2017MNRAS.466.3217Z, 2019ApJ...881L..35S}.

In the bottom three panels of Figure \ref{fig:bpt}, we compare the line ratios from these \oi\ stacks to CLOUDY \citep{2017RMxAA..53..385F} photoionization models following the prescription laid out in \citet{2020ApJ...902L..16J}. The models are based on stellar spectra drawn from BPASS \citep{2017PASA...34...58E,2018MNRAS.479...75S} where each model curve represents a $10^{8.5}$ year-old stellar population with a constant star-formation history. Along each curve of fixed stellar metallicity ($Z_{star}$), we vary the ionization parameter and the nebular metallicity according to the \citet{2020MNRAS.495.4430T} relation: $\rm \log(U)=-1.06 \times [12+\log(O/H)] +5.78$. In both the \nii\ and the \oi\ BPT diagrams, we find that both the $2\leq z\leq 3$ and $1\leq z < 2$ \oi\ stacks agree well with the very sub-solar metallicity ($1.0\times 10^{-5} \lesssim Z_{star} \lesssim 2\times 10^{-3}$) stellar population curves. Since not all of the galaxies in the \oi stacks had wavelength coverage of the \sii $\lambda 6716,6731$ doublet, we do not include them on the \sii\ BPT diagram. Taken together with typical nebular oxygen abundances inferred from the MOSDEF survey \citep{2021ApJ...914...19S,2021MNRAS.506.1237T}, the comparison of these observations with photoionization models supports the picture of harder ionizing spectra from low-metallicity, Fe-poor massive stars driving the line ratios of galaxies at higher redshifts \citep[e.g.,][]{2016ApJ...826..159S,2017ApJ...836..164S,2019ApJ...881L..35S,2020MNRAS.491.1427S,2020MNRAS.495.4430T,2020MNRAS.499.1652T,2021MNRAS.502.2600R,2021MNRAS.505..903C}.

Though the harder ionizing spectrum is fully capable of explaining the enhancement in \oi/H$\alpha$ in these galaxies, it is also possible that shocks and varying contributions of DIG affect the BPT line ratios. With upcoming spectroscopic observations from JWST, an analysis of these effects may become more robust due to a larger sample of galaxies with a wider range of properties, for which we will also have detections of \oi.

\subsection{Nitrogen abundances}
We additionally comment on the nitrogen abundance patterns displayed by our \oiii\ auroral galaxy sample. In the context of the star-forming galaxy population at $z\sim 2$, the nitrogen abundances from the stacks are consistent with empirical predictions based on their average stellar masses. For example, \citet{2022ApJ...925..116S} determined the nitrogen abundances for a sample of 195 $z\sim 2$ star-forming galaxies. Their linear fit to this sample predicts a nitrogen abundance of $\rm 12+\log(N/H)=6.93$ at a stellar mass of $\rm \log(M/M_{\odot})=9.5$ with an intrinsic scatter of 0.33 dex in abundance. Within their respective limits and uncertainties, all three of the stacks as well as GOODS-N-8240 and COSMOS-18812 have nitrogen abundances consistent with this prediction.

\begin{figure}
    \centering
    \includegraphics[width=8cm]{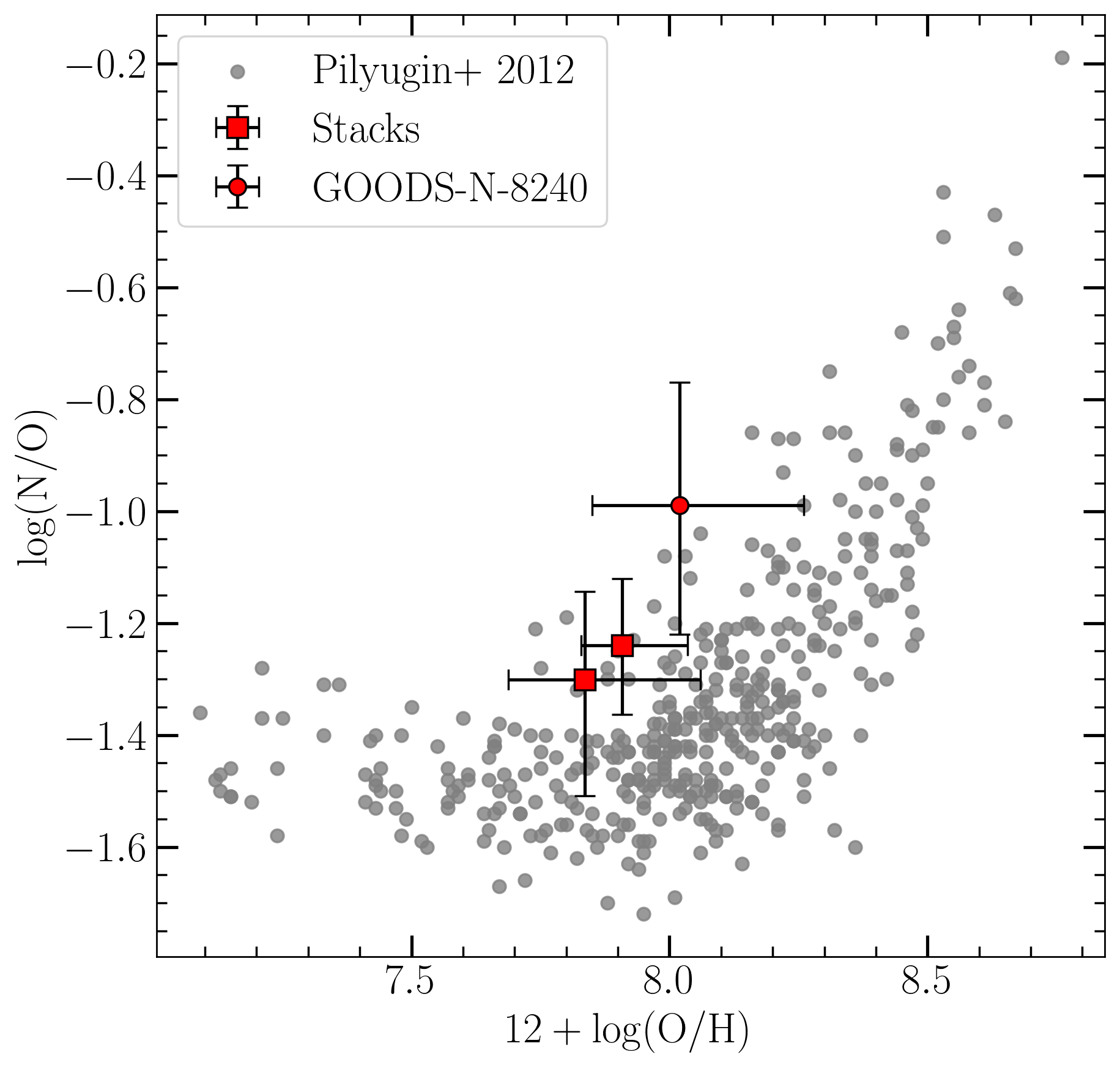}
    \caption{N/O vs. oxygen abundance for stacks 1 and 3 as well as for GOODS-N-8240 compared to local-universe \hii-region measurements from \citet{2012MNRAS.424.2316P}.}
    \label{fig:oh_no}
\end{figure}

As well as analyzing the nitrogen abundance, we discuss the nitrogen to oxygen (N/O) ratio. The N/O ratios of galaxies and \hii regions are often used as a probe of the nucleosynthetic origin of nitrogen where, at low metallicity ($12+\log(\rm O/H) \lesssim 8$), $\log(\rm N/O)$ is fixed at $\sim -$1.5. This is referred to as the ``primary" nitrogen regime since, at low metallicity, the nitrogen yield is tied to those of the $\alpha$ elements \citep{2010ApJ...720.1738P, 2009MNRAS.398..949P, 2006A&A...448..955I}. At higher oxygen abundances, the nitrogen yield increases in proportion to the CNO abundances, and the N/O ratio increases, comprising the ``secondary" nitrogen regime. Since the \oiii\ auroral targets have significantly subsolar oxygen abundances on average, they should fall within the primary nitrogen regime. We plot the N/O ratio vs. oxygen abundance for our sample in Figure \ref{fig:oh_no}.

For stacks 1 and 3, where we have constraints on the N/O ratio, we find that their abundance pattern is consistent with those found in local \hii regions \citep[e.g.,][]{2012MNRAS.424.2316P}. In addition, the N/O ratio for the stacks is similar to those measured within the $z\sim 2.2$ KLEVER sample \citep{2022MNRAS.512.2867H}, though this sample spans higher oxygen abundances $(12+\log{(\rm O/H)} \sim 8.2-8.7)$. However, for GOODS-N-8240, we find that its N/O ratio of $\log(\rm N/O)= -0.99^{+0.22}_{-0.23}$ is slightly enhanced given its oxygen abundance of $12+\log(\rm O/H)=8.02^{+0.24}_{-0.17}$. Several hypotheses have been put forward to explain the abundance pattern of objects with enhanced N/O, one of which appeals to strong winds from Wolf-Rayet stars enriching the ISM \citep[e.g.,][]{1986PASP...98.1005P, 2008A&A...485..657B, 2014ApJ...785..153M}. A more detailed analysis is required to determine the exact source of nitrogen enhancement in this target. 

Another point of interest in studying the N/O ratio is to investigate its effects on trends in the \nii\ BPT diagram across cosmic time. Specifically, if there are significant differences between the N/O vs. O/H ratio at high redshifts compared to low-redshift observations, then an evolving N/O abundance pattern may play an important role in interpreting diagnostic line ratios involving nitrogen and oxygen \citep{2022MNRAS.512.4136C,2022MNRAS.512.2867H}. Because the stacks are consistent with the local N/O vs. O/H relation, there does not appear to be strong evidence for an evolution in the N/O ratio between $z=0$ and $z\sim 2$ based on this sample, though GOODS-N-8240 does represent an outlier in this regard.

\section{Conclusions}

We present an ultra deep rest-optical spectroscopic analysis of several $z>1$, high-excitation galaxies with up to 10 hours of integration time in some bands, and we analyze their excitation properties as well as their oxygen abundances. We selected eight galaxies with strong nebular \oiii\ emission and high predicted electron temperatures to maximize the chance of detecting the \oiii $\lambda 4364$ emission line. We detected \oiii $\lambda 4364$ in two targets, and we chose to stack the eight \oiii-auroral-selected galaxies to observe their general characteristics. Additionally, nine of the galaxies that were not targeted for auroral oxygen emission lines yielded \oi\ detections, enabling the first analysis of high-redshift galaxies in this parameter space. Here are the key conclusions from this work:

\begin{enumerate}
    \item When comparing the oxygen abundnces of the auroral-target stacks and galaxies in the literature on the strong-line indicator diagrams, we find that the stacks from this analysis are qualitatively consistent with the distribution found in the literature in both oxygen abundance and strong-line ratio. In general, it is difficult to say with a small sample size whether the \citet{2020MNRAS.491..944C} or \citet{2018ApJ...859..175B} curves better describe galaxies at $z>1$. In the case of log(R$_{23}$) and log(N$_2$), this may be the case. While the current sample size is limited, these results indicate that stacking analyses are promising.
    
    \item When stacking together the galaxies with \oi\ coverage (both auroral and non-auroral targets), we find that galaxies typically lie at higher \oi/H$\alpha$ at fixed \oiii/H$\beta$ relative to the median locus of local \hii regions. This offset is consistent with photoionization models with low-metallicity ($1.0\times 10^{-5} \lesssim Z_{star} \lesssim 2\times 10^{-3}$) stellar populations, supporting the picture that the line ratios in $z>1$ galaxies are driven by harder ionizing spectra at fixed nebular oxygen abundance.

    \item The N/O abundances of the \oiii\ auroral stacks suggests that the nitrogen enrichment in our galaxy sample at $z\sim 2$ is of primary origin and is consistent with the N/O vs. O/H primary abundance pattern seen in local \hii regions. Though the N/O abundance of GOODS-N-8240 is enhanced given its oxygen abundance, we do not find evidence that the line ratios in our galaxy sample are driven by an evolving N/O ratio with cosmic time.
\end{enumerate}

The results of this analysis demonstrate the limits of 10-m-class ground-based facilities in the realm of nebular metallicity studies of galaxies at cosmic noon. Given that 10 hours of total integration time was still not enough to reach the required depth to consistently detect the \oiii $\lambda 4364$ line in all of the targets, we emphasize the importance of more sensitive facilities such as JWST and future 30-m-class observatories to make advances in this area of study. To date, JWST has already yielded a high number of auroral-line detections out to $z\sim 8$ \citep[e.g.,][]{2023MNRAS.518..425C,2022arXiv221015699W,2023arXiv230112825N,2023arXiv230107072T, 2023arXiv230308149S}. It is already playing an instrumental role in building up the sample of auroral-line measurements at cosmic noon and enabling improvements in strong-line metallicity calibrations at high redshfits.

\section*{acknowledgments}
We acknowledge support from NSF AAG grants 2009313  and 2009085. Support for this work was also provided through the NASA Hubble Fellowship grant \#HST-HF2-51469.001-A awarded by the Space Telescope Science Institute, which is operated by the Association of Universities for Research in Astronomy, Incorporated, under NASA contract NAS5-26555.  We finally wish to extend special thanks to those of Hawaiian ancestry on whose sacred mountain we are privileged to be guests. Without their generous hospitality, the work presented herein would not have been possible.

%

\vspace{5mm}
\facilities{Keck (MOSFIRE)}


\software{Astropy \citep{2018AJ....156..123A}, FAST \citep{2009ApJ...700..221K}, FSPS \citep{2010ascl.soft10043C}, IDL \citep{1993ASPC...52..246L}, PyNeb \citep{2015A&A...573A..42L}, SciPy \citep{2020SciPy-NMeth}
          }



\bibliography{sample631}{}
\bibliographystyle{aasjournal}



\end{document}